\documentclass[aps,amsmath,amssymb,prc,preprint,groupedaddress]{revtex4-2}
\usepackage{amsmath}
\usepackage{bm}
\usepackage{braket}
\usepackage{color}
\usepackage{graphicx}
\usepackage{hyperref}
\usepackage{amssymb}
\usepackage{dcolumn}
\usepackage{multirow}
\usepackage{ulem}

\begin{document}

\title{Quarkyonic Quark-Meson Coupling Model for Nuclear and Neutron Matter}
\author{Koichi Saito}
\email[]{koichi.saito@rs.tus.ac.jp}
\affiliation{Department of Physics and Astronomy, Tokyo University of Science, Noda 278-8510, Japan}
\author{Tsuyoshi Miyatsu}
\email[]{tsuyoshi.miyatsu@ssu.ac.kr}
\affiliation{Department of Physics and OMEG Institute, Soongsil University, Seoul 06978, Republic of Korea}
\author{Myung-Ki Cheoun}
\email[]{cheoun@ssu.ac.kr}
\affiliation{Department of Physics and OMEG Institute, Soongsil University, Seoul 06978, Republic of Korea}
\date{\today}

\begin{abstract}
We unite the dual quarkyonic model with the quark-meson coupling (QMC) model to construct a novel nuclear model based on the quark degrees of freedom, which can cover a wide range of nuclear densities, from low density to the crossover region.  In the model, the relativistic, gaussian quark wavefunction is used to describe the nucleon structure.  We first evaluate the energy density, chemical potential, pressure and sound velocity within the ideal Fermi gas picture.  In this case, those physical quantities are discontinuous or divergent at the quark saturation density, where the quarkyonic phase emerges.  To remove such singular behavior, we next introduce an infrared regulator, and combine the dual quarkyonic model and the QMC model to include the nuclear interaction -- we call it the quarkyonic quark-meson coupling (QQMC) model.  In this model, the quark saturation density depends strongly on the nucleon size.  For example, when $r_p = 0.6\, (0.8)$ fm, where $r_p$ is the root-mean-square radius of the proton, the quark saturation density is about $3.6\,(1.5) \times \rho_0$ in symmetric nuclear matter, where $\rho_0$ is the nuclear saturation density.  Furthermore, the nuclear interaction plays an important role in considering physical quantities quantitatively.  In fact, the QQMC model can produce the sound velocity which is consistent with that inferred from the observed data of several neutron stars.  Furthermore, pressure in symmetric or pure neutron matter deduced from the experiments of heavy-ion collisions at high energy can be explained by the QQMC model as well.  We discuss in detail the formulation for the QQMC model and the physical quantities calculated by the model.  
\end{abstract}

\maketitle

\newpage

\section{Introduction and Preliminaries} \label{sec:introduction} 

For more than a decade, the masses and radii of neutron stars (NSs), together with the constraint on the tidal deformability from the binary neutron-star merger event, GW170817~\cite{LIGOScientific:2017vwq, LIGOScientific:2018hze, LIGOScientific:2018cki}, have been receiving much attention theoretically as well as experimentally, because these observations can provide decisive tests for understanding the equations of state (EoSs) for cold, dense nuclear matter.  In particular, members of the neutron-star population with the heaviest masses are quite important, because, from them, we can obtain information on baryonic and/or exotic compositions in EoSs, which is not accessible through terrestrial experiments.  The observed heavy pulsars are listed as~\cite{NANOGrav:2017wvv}: PSR J1614–2230 ($M_P \sim 1.91 M_\odot$)~\cite{Demorest:2010bx}, PSR J0348+0432 ($M_P \sim 2.01 M_\odot$)~\cite{Antoniadis:2013pzd}, PSR J0740+6620 ($M_P \sim 2.08 M_\odot$)~\cite{Fonseca_2021}, and PSR J0952–0607 ($M_P \sim 2.35 M_\odot$)~\cite{Romani_2022}, where $M_P \, (M_\odot)$ is the mass of pulsar (sun).  Furthermore, the precise measurements of NS radii have been performed with the NICER telescope observation~\cite{Vinciguerra:2023qxq, Choudhury:2024xbk, Salmi:2024aum, Salmi:2024bss, Mauviard_2025}.  

One of the important findings from the NS observations concerns the stiffness of NS matter.  
As the nuclear density, $\rho_N$, increases, conventional nuclear calculations usually lead to a  {\it softening} of EoS for NS matter, because new degrees of freedom like hyperons etc. are opened at high densities.  However, the result from NS observations tells us that the EoS is rapidly {\it stiffened} slightly above the nuclear saturation density, $\rho_0$.  This discrepancy is called {\it the hyperon puzzle}, and a tremendous amount of research in this field has been reported, see, for example, Refs.~\cite{Li2025, 10.3389/fphy.2024.1531475, Brandes_2024, sym17111872}.  

Among those various approaches, the idea of quarkyonic matter, which is originally based on the $1/N_c$ expansion~\cite{MCLERRAN200783} and hadron-quark continuity~\cite{Baym_2018, PhysRevD.102.096017}, seems one of the promising ways to settle the hyperon puzzle.  For review and recent developments, see Refs.~\cite{KOJO2025100088, 831v-8mp4, twcm-zgbq}.  

We first explain the concept of quarkyonic picture very intuitively, which may provide a clue to better understand  quarkyonic matter in the nuclear point of view.  A nucleon at low energy may be approximately viewed as composed of a hard core surrounded by meson clouds.  In the case of an isolated nucleon, the quarks inside a nucleon are confined forever.  However, in nuclear matter, it is {\it not} the case.  In general, the quark wavefunction has a relatively long tail even if it is confined.  In matter, because there are many nucleons, the quark wavefunction can overlap with other quark wavefunctions confined to other nucleons.  When the nuclear density is very dilute, the matter is conventional and is best viewed in terms of nucleons only.  Thus, the probability of such overlap can be ignored and the quark momentum distribution, $f_Q$, is simply characterized by quark compositions inside isolated nucleon.  However, with increasing $\rho_N$, the quarks with low momenta, whose wavefunctions are spread out, can hop from one nucleon to others through the overlap region.  Such quarks can behave as delocalized low-momentum quark modes rather than asymptotically free quarks.  However, they still feel the confinement potential.  It is just like conduction electrons in metal.  Such phase is a novel concept and is called {\it soft deconfinement} in Ref.~\cite{PhysRevD.102.096017}. 

As the density increases further, the overlap region is extended more, and it starts to affect the momentum distributions of quarks, $f_Q(q)$, and nucleons, $f_N(k)$, in matter.  (Hereafter, the letters $q$ and $k$ are used exclusively for quarks and nucleons, respectively.)  Then, the distribution, $f_Q(q)$, increases and eventually reaches the upper bound in Fermi  statistics, i.e. $f_Q(q) = 1$ at $q=0$.  We call this density the {\it quark saturation} density, $\rho_{sat}$.  Above $\rho_{sat}$, the levels with quark momenta between $q=0$ and $q_b$ are fully occupied, i.e. the {\it bulk Fermi sea}, where $q_b$ is the upper momentum of the filled states,  and the matter turns out to be {\it quarkyonic} from ordinary nuclear matter.  In this region, because $f_Q(q)$ is fully occupied, additional quarks with the momenta of $0 \leq q \leq q_b$ are not permitted due to the Pauli blocking at the quark level.  This fact surely reduces the nucleon momentum distribution, $f_N(k)$, with {\it low} momentum, because slow nucleons have relatively more low-momentum quarks which are excluded by the  Pauli blocking.  

The quarkyonic phase thus emerges in the region of soft deconfinement (or in the crossover region), and it is characterized by percolation of quark wavefunction at densities {\it lower} than the threshold density for {\it hard deconfinement} at which the overlap of repulsive hard cores of nucleons is created, that is, it is the usual concept of deconfinement in which the confinement potential melts down.  In other words, the quarkyonic phase may be viewed as a precursor of hard deconfinement.  

Recently, Fujimoto et al.~\cite{PhysRevLett.132.112701, txbp-t8vm} have constructed a novel model of quarkyonic matter, which is called the ideal dual quarkyonic (IdylliQ) model.  See also Ref.~\cite{KOCH2025100025}.  In the IdylliQ model, the nuclear interaction is ignored, and the nucleon is composed of quarks whose wavefunction in momentum space, $\varphi (q)$, is supposed to be a Yukawa-type function 
\begin{equation}
\varphi({\bm q}) = \frac{2\pi^2}{\Lambda^3} \frac{e^{-q/\Lambda}}{q/\Lambda} , 
  \label{eq:yukawa}
\end{equation}
with the typical momentum scale $\Lambda \sim \Lambda_{\rm QCD}$.  Then, collecting quark contributions from each nucleon generates the quark distribution in nuclear matter.  For symmetric nuclear matter, it is described by the sum rule (or duality relation) 
\begin{equation}
f_Q(q) = \int_k \varphi\left( {\bm q} - \frac{\bm k}{N_c} \right) f_N(k) , 
  \label{eq:sumrule}
\end{equation}
where the normalization of $\varphi$ is $\int_k \varphi \equiv \int d^3{\bm k}/(2\pi)^3 \varphi = 1$, $Q$ designates up ($u$) or down ($d$), and $N_c$ is the number of color.  

Actually, in the IdylliQ model, by choosing the Yukawa-type wavefunction, Eq.~(\ref{eq:yukawa}), the sum rule can be transformed into the differential equation, which enables us to solve the IdylliQ model {\it exactly}~\cite{PhysRevLett.132.112701}.  By solving it with appropriate boundary conditions, the nucleon momentum distribution in quarkyonic phase is obtained analytically as a sum of two segments  
\begin{equation}
f_N(k) = \frac{1}{N_c^3} \theta(k_b - k) + \theta(k_s - k) \theta(k - k_b) , 
  \label{eq:postnucleon}
\end{equation}
with $k_b = N_c q_b$.  The first term is the {\it under-occupied} bulk part with the order of $1/N_c^3$, while the second one is the nucleon distribution in the {\it shell} structure at $k_b \leq k \leq k_s$ with $k_s (= N_c q_s)$ the upper bound.  
The quark momentum distribution in quarkyonic phase is calculated analytically as well~\cite{PhysRevLett.132.112701}.  This fact realizes that, above the quark saturation density, pressure, $P$, and the sound velocity, $v_s$, in matter are enhanced considerably.  

At the same time, the IdylliQ model has two excellent advantages: (A) duality holds, (B) a problem with the mismatched Fermi momenta of $u$ and $d$ quarks in quarkyonic phase, which emerges in other quarkyonic models, can be naturally solved.  In particular, duality is important, and it tells us that both of the nuclear density and the energy density of matter can be expressed by either the nucleon or quark momentum distribution.  

In Ref.~\cite{txbp-t8vm}, expanding this idea to the matter including hyperon degrees of freedom, the importance of the quark substructure in neutron matter is investigated. 

In contrast, in the general case where the wavefunction, $\varphi$, is not Yukawa-type and the nuclear interaction is involved, the situation is more complex.  Basically, the momentum distributions, $f_Q(q)$ and $f_N(k)$, should be determined so as to minimize the energy density by optimizing the distributions under the constraints of Fermi statistics and the sum rule, Eq.~(\ref{eq:sumrule}).  It may be a cumbersome task practically.  Furthermore, the sum rule is {\it global}, i.e. $f_Q$ is a functional of $f_N$, which thus makes the model nontrivial.  The difficulty lies in the reconstruction of $f_N(k)$ from a given $f_Q(q)$~\cite{PhysRevD.104.074005}.  At present, a systematic theoretical method for solving this problem has not yet been established.  Recall that, in the IdylliQ model, the momentum distributions can be calculated by solving the differential equation, which is {\it local}.  

So far, in quarkyonic models, the nuclear interaction is not considered.  Thus, it is urgently desirable to consider the effect of interactions among baryons to make the quarkyonic model more realistic.  In Ref.~\cite{PhysRevC.110.025201}, Koch et al. have first performed such an  investigation using the IdylliQ model with the interaction due to $\sigma$ and $\pi$ meson exchanges.  In their calculation, the quarkyonic phase begins below the nuclear saturation density, $\rho_0$.  See also Ref.~\cite{PhysRevC.110.045203, 28sk-9bm3}.  

On the other hand, even below $\rho_0$, using the quark-meson coupling (QMC) model, it has been indicated that subhadronic degrees of freedom are essential to understand the properties of nuclear medium and finite nuclei~\cite{SAITO20071, GUICHON2018262, KREIN2018161}.  The QMC model is originally based on a mean field description of non-overlapping nucleon (or baryon) bags bound by the self-consistent exchange of scalar and vector mesons in the isoscalar and isovector channels~\cite{GUICHON1988235, SAITO19949}.  The model is extended to investigate the properties of finite nuclei, in which, using the Born-Oppenheimer approximation to describe the interacting quark-meson system, one can derive the effective equation of motion for the nucleon (or baryon), as well as the self-consistent equations for the meson mean fields~\cite{GUICHON1996349, SAITO20071, GUICHON2018262}.  In Ref.~\cite{10.1143/PTP.105.373}, using the Miller-Spencer correlation function, the short-range quark-quark correlation is introduced into the QMC model phenomenologically and it is found that the saturation curve for symmetric nuclear matter eventually turns out to be stiff at high density.  Furthermore, by using naive dimensional analysis, it is possible to see that the QMC model can provide remarkably natural coupling constants and hence the model itself is regarded as a natural effective field theory for nuclei~\cite{SAITO1997287}.  

The model can be applied to various finite nuclei, including strange and exotic hypernuclei~\cite{TSUSHIMA19979, GUICHON200866}.   In Ref.~\cite{GUICHON2018262}, 
the extensive application to these topics with a discussion of similarities and differences between the QMC and Skyrme energy density functionals are presented in details.  It is also of great interest that the QMC model predicts a variation of the nucleon form factors in nuclear matter~\cite{LU1998217}, which affects certainly the analysis of electron scattering off nuclei~\cite{PhysRevC.60.068201}, including the polarization-transfer in quasi-elastic $A({\vec e}, e^\prime {\vec p}\,)$ reaction (A1 Collaboration)~\cite{PhysRevC.110.L061302} and the EMC (European Muon Collaboration) effect~\cite{1cf54052d9ae48328912c72f106badaf, ARNEODO1994301, SAITO1994659, CLOET2006210}.  

The QMC model has also been applied to the EoSs for massive NSs and the hyperon puzzle~\cite{Katayama:2012ge, Miyatsu:2011bc, Miyatsu:2015kwa}, and has predicted that, in the EoS for NS, the cascade $\Xi^{-}$ first appears around $\rho_N/\rho_0 \sim 3$ and other hyperons are all suppressed until $\rho_N/\rho_0 \sim 4.5$.  A more phenomenological method to include short range repulsion may be based on the excluded volume effect~\cite{EVERischke1991}.  This method was recently used in the framework of the QMC model~\cite{LEONG2024122928} and the NJL-type model~\cite{sym17040505}. 

Notably, the heart of the QMC model is {\it the scalar  polarizability} of a baryon in matter, i.e. the variation of the quark scalar density inside the nucleon caused by the condensed, scalar mean field, which can naturally describe nuclear saturation and induce many-body forces~\cite{10.1143/PTPS.156.124}.  Therefore, hadrons are polarized in medium even below $\rho_0$, because they have quark substructure.  The medium modification of the nucleon has been examined by lattice calculation (NPLQCD Collaboration)~\cite{PhysRevLett.120.152002} as well. 

The present paper is aimed at unifying the dual quarkyonic picture and the QMC model to construct  a new nuclear model based on the quark degrees of freedom, which is valid from low density to a crossover region (or soft-deconfinement region) where the transition from baryonic to quark matter begins.  Therefore, the new model includes the effect of Pauli blocking at the quark level as well as the scalar polarizability of nucleon in medium.  

In the IdylliQ model, the single quark momentum distribution in a nucleon is chosen to be the Yukawa-type function, Eq.~(\ref{eq:yukawa}), specifically, while in the QMC model the quark wavefunction is usually given by the bag model.  Therefore, first of all, it is necessary to prepare a unified setup for the quark wavefunction.  In this paper, we adopt a gaussian function as the quark wavefunction in a nucleon, which is given by the relativistic confinement potential of the scalar-vector harmonic oscillator (HO) type.  The QMC model with the  relativistic HO potential is sometimes referred to as the quark mean field (QMF)  model~\cite{PhysRevC.58.3749, PhysRevC.61.045205} or the modified quark-meson coupling (MQMC) model~\cite{PhysRevC.88.015206}.  This quark model has the useful feature that all quantities of interest can be basically calculated analytically.  

We first consider a dual quarkyonic model  in ideal Fermi gas.  It is thus identical to the IdylliQ model, but the wavefunction is gaussian, instead of the Yukawa-type one.  We call this the gaussian quarkyonic (GQ) model.  This indicates that, since the gaussian wavefunction is relativistic, due to the lower component, the quark saturation density turns out to be higher than in the case where a non-relativistic wavefunction is used.  In the GQ model, we can find that $\rho_{sat}$ is never lower than $\rho_0$.  

In the GQ model, since it is naive, we encounter the following three troubles.  As previously stated, because the quark wavefunction is not Yukawa-type, it is unavailable to use the differential equation derived in the IdylliQ model, and it is thus very hard to find the exact momentum distributions of quarks, $f_Q$, and nucleons, $f_N$, simultaneously.  This is the first trouble.  However, from the discussion in Refs.~\cite{KOJO2025100088, PhysRevD.104.074005}, we can deem that Eq.~(\ref{eq:postnucleon}) is still a good guess even in the GQ model.  The second one is how the two characteristic momenta in quarkyonic phase, $k_b$ and $k_s$, should be determined.  To settle this issue, we choose the same boundary condition as in the IdylliQ model, because it should be satisfied in any case, regardless of the quark wavefunction.  Using those assumptions, we can estimate the nucleon momentum distribution, $f_N$.  Once $f_N$ is obtained, due to duality, we can calculate various physical quantities (energy density, chemical potential, pressure, sound velocity) over the wide range of $\rho_N$.  However, as in the IdylliQ model, they show singular behavior at the quark saturation density, which should be remedied.  This is the last problem.  To control it, we introduce an infrared regulator, which may be regarded as an effective parametrization of the smearing of the Fermi surface caused by quark-exchange and gluonic interactions among neighboring nucleons.  Then, we can obtain the physical quantities in symmetric nuclear and pure neutron matter, which behave continuously or smoothly at $\rho_{sat}$.  

Next, we construct a dual quarkyonic model including the nuclear interaction, which is called the quarkyonic quark-meson coupling (QQMC) model, by unifying the GQ model and the QMC model.  As explained previously, because the QMC model can be viewed as an extension of Quantum Hadrodynamics (QHD) to include the quark degrees of freedom, the physical quantities like the nuclear density, energy density, etc. can be expressed by those of a relativistic Fermi gas of nucleons with in-medium mass, $M_N^\ast$.  Furthermore, both quark and nucleon degrees of freedom couple to the common mean scalar and vector fields.  Thus, we assume that duality still holds in the QMC framework, and we construct the QQMC model in terms of the nucleon degrees of freedom only.    

Here, it should be notable that the quark saturation density in matter, $\rho_{sat}^\ast$, becomes lower than the value in the GQ model, because a nucleon in matter is swelled by the attractive, scalar field.  However, we can confirm that $\rho_{sat}^\ast$ is still larger than $\rho_0$.  This is important and indicates that the QQMC model below $\rho_{sat}^\ast$ is  entirely consistent with the present status of nuclear experiments at low energy.  It is, however, remarkable that, in symmetric nuclear matter, $\rho_{sat}^\ast$ is not so far from $\rho_0$, i.e. $\rho_{sat}^\ast \simeq 1.5 \rho_0$ when the nucleon radius in free space is $0.8$ fm.  

Finally, using the QQMC model, we again calculate the energy density, chemical potential, pressure, and sound velocity over the range of $0 \leq \rho_N/\rho_0 < 5 - 10$.  In particular, the QQMC model can provide the sound velocity which is consistent with that inferred from the observed neutron-star data by using the neural network model or Bayesian inference analysis.  Furthermore, the pressure calculated by the QQMC model lies within the range deduced from the experimental flow data in heavy ion collisions at high energy and the kaon production data.  As the quark saturation density relies on the size of nucleon, $r_p$, various physical quantities also depend on it accordingly.  From the present consideration, as expected, the choice of $r_p = 0.6 - 0.8$ fm seems most suitable for describing dense nuclear matter in the QQMC model.  The QQMC model provides a unified framework in which both quark-level Pauli blocking and medium modification of nucleon structure are incorporated simultaneously. This combination is essential for describing dense matter in the crossover region between hadronic and quark degrees of freedom.

Here is an outline.  In Sec.~\ref{sec:freecase}, using the gaussian quark wavefunction, we first study the quarkyonic matter in ideal Fermi gas.  We discuss the feature of the gaussian quarkyonic model, and calculate the energy density, chemical potential, pressure and sound velocity in symmetric or pure neutron matter.  Furthermore, we consider a minimal correction to the gaussian quarkyonic model to remove the singular behavior at the quark saturation density.   In Sec.~\ref{sec:qqmc}, we briefly review the quark-meson coupling model with a harmonic oscillator potential, and build a new, quark-based nuclear model, in which we incorporate the QMC model and the gaussian quarkyonic model.  We then discuss the results of the present model in detail.  Lastly, in Sec.~\ref{sec:concandsum}, we give our summary and discussion.  In Appendix~\ref{app:sommerfeld}, we apply Sommerfeld expansion to the gaussian quarkyonic model.

\newpage
\section{Quarkyonic matter in ideal Fermi gas} \label{sec:freecase}

In this section, we study a quarkyonic model with the relativistic, gaussian quark wavefunction.  We here neglect the nuclear interaction, and consider the effect of Pauli blocking at the quark level only.  We call it the gaussian quarkyonic (GQ) model.  Using the GQ model, we calculate several physical quantities in symmetric nuclear matter (SNM) and in  pure neutron matter (PNM).  

\subsection{Relativistic quark model of nucleon} \label{subsec:relquarkmod}

We want to use a gaussian function as the quark wavefunction in a nucleon, which is given by the relativistic confinement potential of the scalar-vector harmonic oscillator (HO) type 
\begin{equation}
  U(r) = \frac{c}{2} (1+\gamma_0) r^2 , 
  \label{eq:conpot}
\end{equation}
with the strength parameter $c$.  It is well known that the Dirac equation with the HO potential, Eq.~(\ref{eq:conpot}), can be solved analytically~\cite{ferreira771, ferreira772, PhysRevD.110.113001}.  In free space, the lowest-state solution for a iso-symmetric ($Q=u=d$) quark is given by
\begin{align}
\psi_{Q}({\bm r}) 
  & = \frac{1}{\pi^{3/4} a^{3/2}} \sqrt{\frac{2 \lambda^2 a^2}{2\lambda^2 a^2+3}}
  \begin{pmatrix}
1 \\
i {\vec \sigma} \cdot {\hat r} \frac{1}{\lambda a} \left( \frac{r}{a} \right) 
\end{pmatrix}
  e^{-r^2/2a^2} \chi ,  \label{eq:sol} \\
  \lambda & = \epsilon + m, \ \ \  a^2 = \frac{1}{\sqrt{c\lambda}} = \frac{3}{\epsilon^2-m^2} ,    \label{eq:norm}
\end{align}
where $m$ is the constituent quark mass, $a$ is the width and $\chi$ is the quark spinor.  Here, the single-particle quark energy, $\epsilon$, is determined by
\begin{equation}
  \sqrt{\epsilon + m} (\epsilon - m) = 3 \sqrt{c} .
  \label{eq:qenergy}
\end{equation}
Note that, in the limit $\lambda \to \infty$, this wavefunction turns out to be the nonrelativistic (NR) one.  

The zeroth-order energy of the nucleon, $E_N^0$, is then simply given by a sum of the quark energies, $E_N^0 = 3 \epsilon$.  Now we should take into account some corrections to the nucleon energy such as the spin correlations, $E^{spin}_N$, due to the quark-gluon and quark-pion interactions~\cite{PhysRevD.33.1925, PhysRevC.108.025809}, and the center of mass (c.m.) correction, $E^{c.m.}_N$.  Here we do not calculate the spin correlation explicitly, but we assume that it can be treated as a constant parameter which is fixed so as to reproduce the nucleon mass~\cite{PhysRevC.58.3749, PhysRevC.61.045205}.  The c.m. correction to the spurious motion can be calculated analytically~\cite{PhysRevD.110.113001}  
\begin{equation}
  E^{c.m.}_{N}
  =\frac{77 \epsilon + 31m}{3a^2(3\epsilon + m)^2} .   \label{eq:cmcorrection} 
\end{equation}
The nucleon mass in vacuum is thus given by 
\begin{equation}
  M_N = E_N^0 + E^{spin}_N - E^{c.m.}_N , 
  \label{eq:bmass}
\end{equation}
which is equal to the proton or neutron mass ($M_N=M_p=M_n$), because any breaking of isospin symmetry is not included.  
We also find the root-mean-square charge radius of proton including the c.m. correction as~\cite{PhysRevD.110.113001}   
\begin{equation}
  r_p \equiv \langle r^2 \rangle_{p}^{1/2} = \sqrt{\frac{11\epsilon + m}{\left(3\epsilon + m \right)\left(\epsilon^2-m^2\right)}}  . \label{eq:radiusN} 
\end{equation}

In this quark model,  we have three parameters, $c$, $m$, $E_N^{spin}$.  We then choose the quark mass as follows: case 1, $m = 250$ MeV; case 2, $m = 300$ MeV; case 3, $m = 350$ MeV.  The remaining two parameters, $c$ and $E_N^{spin}$, are determined so as to reproduce the nucleon mass, $M_{N}=939$ MeV, and the proton charge radius, $r_p = 0.6, 0.7, 0.8, 0.9$ fm.  
Note that the experimental value of the charge radius of proton is $0.8409 \pm 0.0004$ fm~\cite{PhysRevD.110.030001}, while $r_p$ in the present calculation may correspond to the charge radius of valence quarks.
In Table~\ref{tab:parameters}, the parameters and the quark energies are summarized.  
\begin{table}[t!]
  \caption{\label{tab:parameters}
   Parameters in the quark model and the quark saturation density, $\rho_{sat}$.  The nuclear saturation density is chosen to be $\rho_{0}=0.15$ fm$^{-3}$.  The quark saturation density is listed in the last two columns (see Section\,\ref{subsubsec:atsaturation}). 
  }
  \begin{ruledtabular}
    \begin{tabular}{ccccccccccc}
      \multirow{2}{*}{Case} & $m$\footnote{Input}                  & $r_{p}^{\textrm a}$
       & $a$   & $c$         & $\epsilon$ & $E_{N}^{0}$ & $E_{N}^{spin}$ & $E_{N}^{c.m.}$ & \multicolumn{2}{c}{$\rho_{\textrm{sat}}/\rho_0$} \\ 
      \cline{2-2}\cline{3-4}\cline{5-5}\cline{6-9}\cline{10-11}
      \                     & (MeV)                & \multicolumn{2}{c}{(fm)}           & (fm$^{-3}$) & \multicolumn{4}{c}{(MeV)}                                  & \textrm{SNM}       & \textrm{PNM}                                           
\\
      \colrule
      \multirow{4}{*}{1}    & \multirow{4}{*}{250} &                        0.6 & 0.567 & 2.121       & 653.0      & 1959.1      & $-539.2$       & 480.9          & 4.372                               & 3.370                                \\
      \                     &                    \ &                        0.7 & 0.665 & 1.230       & 571.8      & 1715.3      & $-382.5$       & 393.8          & 2.648                                & 2.042                                \\
      \                     &                    \ &                        0.8 & 0.764 & 0.761       & 512.7      & 1538.2      & $-270.3$       & 328.9          & 1.714                                & 1.321                                \\
      \                     &                    \ &                        0.9 & 0.863 & 0.496       & 468.4      & 1405.2      & $-187.4$       & 278.8          & 1.167                                & 0.900                                \\
      \cline{1-11}
      \multirow{4}{*}{2}    & \multirow{4}{*}{300} &                        0.6 & 0.570 & 1.922       & 670.2      & 2010.5      & $-616.3$       & 455.2          & 4.175                                & 3.219                                \\
      \                     &                    \ &                        0.7 & 0.669 & 1.102       & 592.3      & 1776.8      & $-468.9$       & 368.8          & 2.526                                & 1.947                                \\
      \                     &                    \ &                        0.8 & 0.769 & 0.675       & 536.3      & 1608.8      & $-365.0$       & 304.8          & 1.634                                & 1.260                                \\
      \                     &                    \ &                        0.9 & 0.869 & 0.436       & 494.7      & 1484.0      & $-289.1$       & 255.9          & 1.114                                & 0.859                                \\
      \cline{1-11}
      \multirow{4}{*}{3}    & \multirow{4}{*}{350} &                        0.6 & 0.574 & 1.750       & 691.0      & 2072.9      & $-703.6$       & 430.3          & 4.010                                & 3.092                                \\
      \                     &                    \ &                        0.7 & 0.673 & 0.993       & 616.6      & 1849.7      & $-565.6$       & 345.1          & 2.426                                & 1.871                                \\
      \                     &                    \ &                        0.8 & 0.774 & 0.603       & 563.7      & 1691.1      & $-469.6$       & 282.5          & 1.572                               & 1.212                                \\
      \                     &                    \ &                        0.9 & 0.874 & 0.387       & 524.8      & 1574.5      & $-400.3$       & 235.2          & 1.073                                & 0.828                                \\
    \end{tabular}
  \end{ruledtabular}
\end{table}

Using the gaussian quark wavefunction, Eq.~(\ref{eq:sol}), the single quark momentum distribution in a single nucleon is calculated as  
\begin{equation}
\varphi({\bm q}) = \pi^{3/2} a^3 \left( \frac{16 \lambda^2 a^2}{2 \lambda^2 a^2 +3} \right) 
\left( 1+ \frac{q^2}{\lambda^2} \right) e^{- a^2 q^2}  ,  \label{eq:qmom} 
\end{equation}
where the second term in the right-hand side comes from the lower component of the quark wavefunction.  This satisfies the normalization condition, $\int_q \varphi({\bm q}) = 1$.  Through the present paper,  we use this single quark momentum distribution.  

\clearpage

\subsection{Gaussian quarkyonic model} \label{subsec:gaussianmodel}

In the dual quarkyonic model, the nuclear density, $\rho_N$, and the energy density, $\epsilon_N$, are expressed by either the nucleon, $f_N(k)$, or quark momentum distribution, $f_Q(q)$: 
\begin{align}
&\rho_N = \gamma \int_k f_N(k) = \gamma \int_q f_Q(q)  ,  \label{eq:dualdensity}  \\
&\epsilon_N[f_N] = \epsilon_Q[f_Q] , \ \ \epsilon_N[f_N] = \gamma \int_k E_N(k) f_N(k) , \ \ \epsilon_Q[f_Q] = \gamma \int_q E_Q(q) [N_c f_Q(q)]  ,     \label{eq:dualenergy}
\end{align}
with $\gamma = 4 \, (2)$ the spin-isospin degeneracy for SNM (PNM), $E_N$ the nucleon energy, and $E_Q$ the quark energy.  This duality is ensured by the sum rule, Eq.~(\ref{eq:sumrule}).  
On the other hand, as explained in Sec.~\ref{sec:introduction}, it is difficult to calculate the exact  momentum distributions, $f_N(k)$ and $f_Q(q)$, because the quark wavefunction is chosen to be gaussian.  Instead, we shall determine the momentum distributions as in the following subsections.  

\subsubsection{Nuclear density} \label{subsubsec:nucleardensity}

The nuclear density, $\rho_{N}$, in SNM or PNM is expressed by 
\begin{equation}
\rho_N = \frac{\gamma}{2\pi^2} \int dk\, k^2 f_N(k)  .   \label{eq:nucleonden} 
\end{equation}
Here, we note that $f_N(k) = f_p(k) = f_n(k)$ for SNM, while $f_N(k) = f_n(k), \, f_p(k)=0$ for PNM, where $f_{p \, (n)}(k)$ is the momentum distribution for protons (neutrons).  
The nuclear density is also described in terms of the quark momentum distribution $f_Q$ as in Eq.~(\ref{eq:dualdensity}). 

\subsubsection{Presaturation region} \label{subsubsec:presaturation}

Because, below the quark saturation density, the nuclear matter is normal, 
the momentum distribution in Eq.~(\ref{eq:nucleonden}) is simply given by the usual Fermi distribution 
\begin{equation}
f_N(k) = \theta(k_s - k)  ,   \label{eq:prenucleonden}  
\end{equation}
where $k_s = k_F$ ($k_F$ the Fermi momentum).  

In SNM, the dual momentum-space distribution for iso-symmetric light quarks with a fixed  color is expressed by the sum rule (see Eq.~(\ref{eq:sumrule})) 
\begin{equation}
f_{Q}(q, k_s) = \int_k \varphi\left( {\bm q} - \frac{\bm k}{N_c} \right) f_N(k) ,   \label{eq:snmsumrule}  
\end{equation}
which is less than unity in the presaturation region.  Then, we can calculate $f_Q$ analytically 
\begin{align}
f_Q(q, k_s) &= \frac{N_c^3}{\sqrt{\pi}} \bigg[ {\rm Er}({\bar q}) + {\rm Er}(-{\bar q}) - {\rm Er}({\bar k}_s+{\bar q}) - {\rm Er}({\bar k}_s-{\bar q})   \nonumber \\
&+ \frac{1}{2{\bar q}} \left( 1+ \frac{2{\bar k}_s^2+1}{2{\bar \lambda}^2+3} \right) \left( e^{-({\bar k}_s+{\bar q})^2} - e^{-({\bar k}_s-{\bar q})^2} \right)  
\left. + \frac{{\bar k}_s}{2{\bar \lambda}^2+3} \left( e^{-({\bar k}_s+{\bar q})^2} + e^{-({\bar k}_s-{\bar q})^2} \right) \right]     \label{eq:fq1}
\end{align}
where ${\bar q} \equiv aq$, ${\bar \lambda} \equiv a\lambda$, and ${\bar k} \equiv ak/N_c$ are dimensionless variables.  We here define the error function as 
\begin{equation}
{\rm Er}(x) \equiv \int_x^\infty dt \, e^{-t^2} .   \label{rq:error}
\end{equation}

In contrast, in PNM, there are only neutrons ($udd$) in matter, and so the saturation of $d$ quark first occurs because the $u$ quark levels are half-occupied compared with the $d$ quark levels.  
Therefore, the sum rule for PNM reads~\cite{txbp-t8vm}
\begin{equation}
f_d(q, k_s) = B_d^n \int_k \varphi\left( {\bm q} - \frac{\bm k}{N_c} \right) f_n(k)  ,   \label{eq:pnmsumrule} 
\end{equation}
where the factor, $B_d^n=2/3$, is the baryon number of $d$ quarks in a neutron. 
It can be calculated analytically as well.

\subsubsection{At quark saturation density} \label{subsubsec:atsaturation}

At quark saturation density, the quark momentum distribution first reaches unity at $q=0$.  In SNM, the nucleon momentum $k_{sat}$ at $\rho_{sat}$ is given by the saturation condition 
\begin{equation}
g_Q(k_{sat}) = \frac{N_c^3}{\sqrt{\pi}} \left[ \sqrt{\pi} - 2{\rm Er}({\bar k}_{sat}) 
- 2 {\bar k}_{sat} \left( 1 + \frac{2{\bar k}_{sat}^2}{2{\bar \lambda}^2+3} \right) e^{-{\bar k}_{sat}^2} \right] = 1   ,  \label{eq:gQ0}
\end{equation}
where we define $g_Q(k) \equiv f_Q(0, k)$.  
Similarly, in PNM, the neutron momentum $k_{sat}$ at $\rho_{sat}$ is calculated by 
\begin{equation}
g_d(k_{sat}) = 1   ,  \label{eq:gd0}
\end{equation}
with $g_d(k) \equiv f_d(0, k)$. 

In the last two columns of Table~\ref{tab:parameters}, we show the quark saturation densities in SNM and in PNM.  
In the NR limit, we find that, when $r_p = 0.8$ fm, $\rho_{sat}/\rho_0 \simeq 1.27$ in SNM.  Here, we take $\rho_0 = 0.15$ fm$^{-3}$.  In contrast, when $m = 300$ MeV and $r_p = 0.8$ fm, the present relativistic calculation predicts $\rho_{sat}/\rho_0 \simeq 1.63$, which is larger than in the NR result.  This discrepancy is caused by the lower component of the relativistic wavefunction, which enhances the higher momentum part in the quark wavefunction, $\varphi({\bm q})$.  Furthermore, the quark saturation density in SNM is larger than that in PNM by about $30 \%$.  Such tendency is seen in the IdylliQ model as well, although the difference in the IdylliQ model is tiny.  
We note that, in Ref.~\cite{{PhysRevC.110.045203}}, using the transverse momentum densities for the valence and sea quarks in a nucleon, $\rho_{sat}/\rho_0$ is estimated to be about $1.13$. 

The quark saturation density strongly depends on the size of nucleon.  
This strong  dependence originates from the width of the quark wavefunction, which controls the population of low-momentum quark modes and thus determines the onset of quark-level Pauli blocking.  On the other hand, the dependence on the quark mass is approximately linear 
and weak, i.e. fixing $r_p$, the difference between $\rho_{sat}$ in the case 1 and that in case 3 is less than $10\%$.  Hence, hereafter, we consider the case 2, i.e. $m=300$ MeV, only.

\subsubsection{Postsaturation region} \label{subsubsec:postsaturation}

Above the quark saturation density, the low momentum levels of quarks are fully occupied, where the quarks behave like Fermi gas but still feel the confinement force, i.e. {\it soft deconfinement}.  Such region is denoted by $0 \leq q \leq q_b$, and we posit $f_{Q \, (d)}^{<q_b}(q) = \theta(q_b - q)$ for SNM (PNM).  
Correspondingly, the dual nucleon momentum distribution should be strongly suppressed by the quark Pauli blocking.  As explained in Sec.~\ref{sec:introduction}, Eq.~(\ref{eq:postnucleon}) may be still a good guess even in the GQ model~\cite{PhysRevD.104.074005}.  We thus assume that $f_{N}^{<k_b}(k) = \frac{1}{N_c^3} \theta(k_b - k)$ for SNM and $f_{n}^{<k_b}(k) = \frac{1}{B_d^n N_c^3} \theta(k_b - k)$ for PNM, which are the under-occupied bulk part ($0 \leq k \leq k_b = N_c q_b$).  Here, the suppression factor, $1/N_c^3$, can be calculated by Eq.~(\ref{eq:sumrule}) with the wavefunction in the limit $a \to \infty$, because quarks in the bulk Fermi sea are free.  Alternatively, this factor can be found through the discussion of phase space density~\cite{PhysRevD.104.074005}.  

On the other hand, for the higher momentum states of quarks, i.e. $q > q_b$, the levels are not fully occupied, and thus the quark momentum distribution is expected to be expressed by $\theta(q - q_b) f_{Q \, (d)}^{>q_b}(q)$, where $f_{Q\, (d)}^{>q_b}(q=q_b)=1$ and $f_{Q\, (d)}^{>q_b}(q)$ is a decreasing function from unity.  Thus, we assume that $f_{Q\, (d)}^{>q_b}(q) = f_{Q\, (d)}(q-q_b, k_{sat})$, as discussed in Ref.~\cite{PhysRevD.104.074005}.  
In this region, because the Pauli blocking at the quark level does not work, the nucleon momentum distribution is not suppressed and it should form the ordinary shell structure, i.e. $f_{N\, (n)}^{>k_b}(k) = \theta(k_s - k) \theta(k - k_b)$, where $k_s$ is the upper bound given by $\rho_{N}$.  
It is thus expected that the nucleon momentum distributions for SNM and PNM can be expressed respectively 
\begin{align}
f_N(k) &= \frac{1}{N_c^3} \theta(k_b - k) + \theta(k_s - k) \theta(k - k_b)  ,   \label{eq:nucleondist1}    \\
f_n(k) &= \frac{1}{B_d^n N_c^3} \theta(k_b - k) + \theta(k_s - k) \theta(k - k_b)  .   \label{eq:nucleondist2}  
\end{align}

Now, the problem is how we should determine the momenta at the boundary, $k_b$, and at the upper bound, $k_s$.   Substituting the momentum distributions, Eqs.~(\ref{eq:nucleondist1}) and (\ref{eq:nucleondist2}), into the sum rules, Eqs.~(\ref{eq:snmsumrule}) and (\ref{eq:pnmsumrule}), respectively,  we can calculate the quark momentum distributions in the postsaturation region.  Then, at $q=0$, their values should be {\it unity}: 
\begin{align}
g_Q(k_s) &- \beta g_Q(k_b) = 1   \ \ \ \ \  {\rm for \ SNM},   \label{eq:determination1}    \\
g_d(k_s) &- \beta^{\prime} g_d(k_b) = 1   \ \ \ \ \ {\rm for \ PNM}  ,   \label{eq:determination2}  
\end{align}
with $\beta = 1- 1/N_c^3$ and $\beta^{\prime} = 1- 1/(B_d^nN_c^3)$.  Note that, when $k_s=k_{sat}$, $k_b$ vanishes.  
The boundary conditions, Eqs.~(\ref{eq:determination1}) and (\ref{eq:determination2}), are the {\it necessary} condition, and they should hold in any case, regardless of the quark wavefunction, $\varphi({\bm q})$.  Furthermore, they are identical to the condition that sets the relation between $k_b$ and $k_s$ in the IdylliQ model~\cite{PhysRevLett.132.112701}.  Thus, in this paper we posit these boundary conditions to determine $k_b$ and $k_s$.  
  
\begin{figure}[t!]
  \includegraphics[width=12.0cm,keepaspectratio,clip]{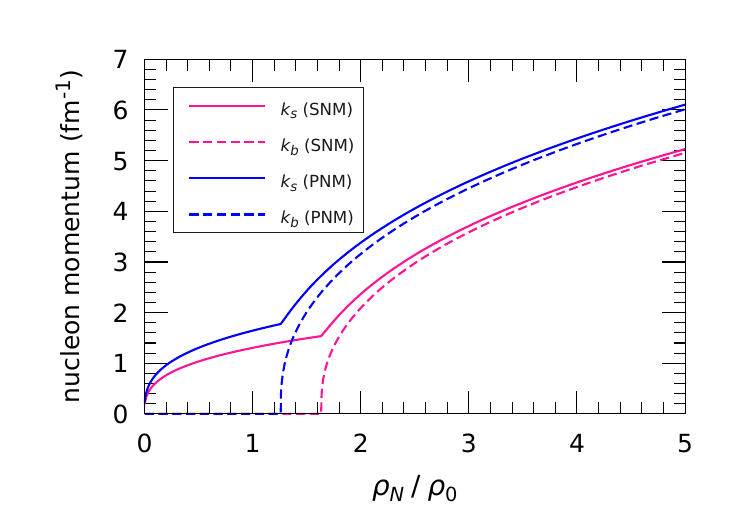}
  \caption{\label{fig:kbks} Density dependence of $k_{s}$ and $k_{b}$ in the case of $r_p=0.8$ fm. 
  }
\end{figure}
\begin{figure}[t!]
  \includegraphics[width=15.0cm,keepaspectratio,clip]{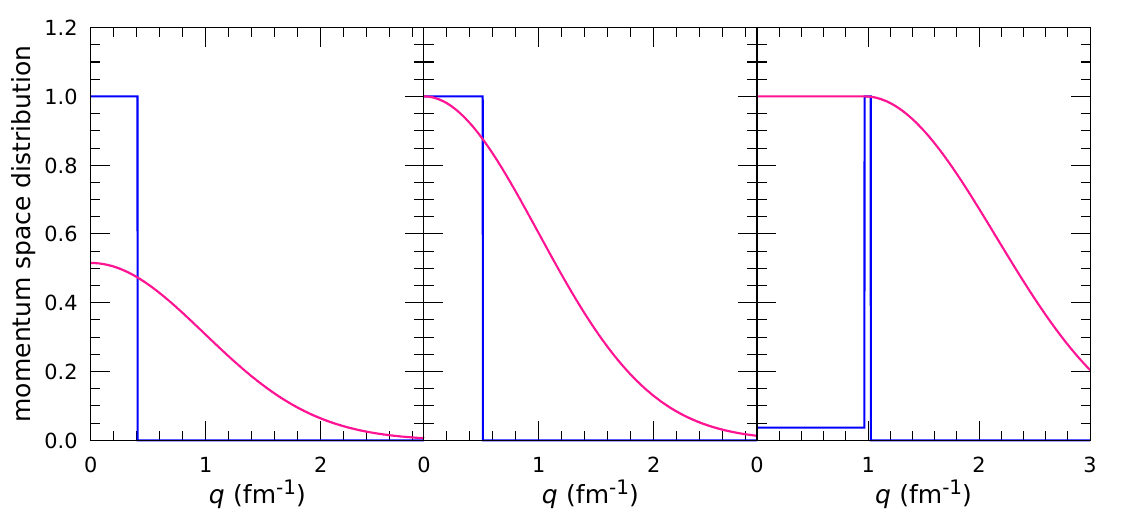}
  \caption{\label{fig:dist08} Quark and nucleon momentum distributions in symmetric nuclear matter.  We use the GQ model with $r_p=0.8$ fm.   The left (middle) [right] panel is for $\rho_N/\rho_{sat} = 0.5\, (1.0)\, [1.5]$.  The distributions, $f_Q(q) \, (f_N(k)$ with $k=qN_c$), are denoted by the red (blue) lines. 
  }
\end{figure}
In Fig.~\ref{fig:kbks}, we show the density dependence of $k_b$ and $k_s$ in SNM or in PNM.  The rapid rise of the momenta, $k_b$ and $k_s$, occurs beyond $\rho_{sat}$, which leads to singular behavior of thermodynamic quantities.  Furthermore, in Fig.~\ref{fig:dist08}, we illustrate the quark and nucleon momentum distributions, $f_Q(q)$ and $f_N(k)$, in SNM.  

\clearpage

\subsection{Initial results} \label{subsec:results1}

Using the GQ model, we evaluate the energy density, chemical potential, pressure and sound velocity from the viewpoint of nucleon matter.  

\subsubsection{Symmetric nuclear matter} \label{subsubsec:snmresults1}

First, the nuclear density, Eq.~(\ref{eq:nucleonden}), reads 
\begin{align}
\rho_N^{below} & = \frac{2}{3\pi^2} k_s^3  \ \ \ \ \ \ \  (k_s = k_F),   \label{eq:rhobelow}    \\
\rho_N^{above} & = \frac{2}{3\pi^2}  ( k_s^3 - \beta k_b^3 )   ,   \label{eq:rhoabove}  
\end{align}
where the superscripts (below, above) indicate ($\rho_N < \rho_{sat}$, $\rho_N > \rho_{sat}$), respectively.  Because, in the limit $\rho_N \to \rho_{sat}$, $k_b$ approaches zero, the density is continuous at $\rho_{sat}$.  The derivative of the density with respect to $k_s$ is then given by 
\begin{align}
\frac{\partial \rho_N^{below}}{\partial k_s} & = \frac{2}{\pi^2} k_s^2  ,   \label{eq:derrhobelow}    \\
\frac{\partial \rho_N^{above}}{\partial k_s}  & = \frac{2}{\pi^2} k_s^2 \left[ 1 - \beta  \frac{k_b^2}{k_s^2}\,  \frac{\partial k_b}{\partial k_s} \right] = \frac{2}{\pi^2} k_s^2 \left[ 1 -  \frac{k_b^2}{k_s^2} \, \frac{\phi(k_s)}{\phi(k_b)} \right] .  
\label{eq:derrhoabove}  
\end{align}
Here, the derivative of $k_b$ with respect to $k_s$ is calculated by the boundary condition, Eq.(\ref{eq:determination1}), as 
\begin{equation}
\frac{\partial k_b}{\partial k_s} =  \frac{1}{\beta} \frac{\phi(k_s)}{\phi(k_b)} = \frac{1}{\beta} \frac{r_1(k_s){\bar k}_s^2}{r_1(k_b){\bar k}_b^2} \, e^{- {\bar k}_s^2 + {\bar k}_b^2}   ,   \label{eq:derkbks}
\end{equation}
where we define 
\begin{equation}
\phi(k) \equiv \frac{\partial}{\partial {\bar k}} g_Q(k) = \frac{N_c^3}{\sqrt{\pi}} [ 4 r_1(k) {\bar k}^2] e^{- {\bar k}^2}  ,   \label{eq:dergq}
\end{equation}
with
\begin{equation}
r_n(k) = 1 - \frac{(2n+1) - 2 {\bar k}^2}{2 {\bar \lambda}^2 + 3}  .   \label{eq:ration}
\end{equation}
Note that $g_Q(k)$ and $\phi(k)$ are dimensionless. 

In Eq.~(\ref{eq:derrhoabove}), the second term in the right-hand side has the factor $k_b^2$ coming from the Jacobian, but it is completely canceled by the factor $k_b^{-2}$ in Eq.~(\ref{eq:derkbks}).  This is an outstanding characteristic in the case of gaussian quark wavefunction.  Thus, unfortunately, for chemical potential, pressure or sound velocity, the continuity at $\rho_{sat}$ does {\it not} hold.  

We next calculate the energy density, $\epsilon_N$.  We find 
\begin{align}
\epsilon_N^{below} & = \frac{2}{\pi^2} \int_0^{k_s} dk \, k^2 E_N(k) = \frac{2}{\pi^2}  I(M_N, k_s) ,  \label{eq:energybelow}    \\
\epsilon_N^{above} & = \frac{2}{\pi^2}  \left( \int_0^{k_s} - \beta \int_0^{k_b} \right) dk \, k^2 E_N(k) = \frac{2}{\pi^2}  [ I(M_N, k_s) - \beta I(M_N, k_b) ] ,   \label{eq:energyabove}  
\end{align}
with $E_N(k) = \sqrt{M_N^2 + k^2}$ and 
\begin{align}
I(M, x) &= \int_0^x dt \, t^2 \sqrt{M^2 + t^2}    \nonumber \\
&= \frac{1}{8} \left[ x (M^2 + 2x^2) \sqrt{M^2 + x^2} + M^4 \log \left( \frac{M}{x + \sqrt{M^2 + x^2}} \right) \right] .   \label{eq:integ1}
\end{align}
Therefore, the energy density is continuous at $\rho_{sat}$.  However, the first derivatives are given by  
\begin{align}
\frac{\partial \epsilon_N^{below}}{\partial k_s} & = \frac{2}{\pi^2} k_s^2 E_N(k_s) ,   \label{eq:derenergybelow}    \\
\frac{\partial \epsilon_N^{above}}{\partial k_s}  & = \frac{2}{\pi^2} k_s^2 E_N(k_s) \left[ 1 - \beta \frac{k_b^2E_N(k_b)}{k_s^2E_N(k_s)}\,  \frac{\partial k_b}{\partial k_s} \right]= \frac{2}{\pi^2} k_s^2 E_N(k_s) \left[ 1 - \frac{k_b^2 E_N(k_b)}{k_s^2E_N(k_s)} \, \frac{\phi(k_s)}{\phi(k_b)}  \right]  . \label{eq:derenergyabove}  
\end{align}
So, the derivative has a gap at $\rho_{sat}$.  

The chemical potential, $\mu_N = \partial \epsilon_N / \partial \rho_N$, is similarly evaluated as
\begin{align}
\mu_N^{below} & = E_N(k_s)  ,  \ \ \ \ \   \frac{\partial \mu_N^{below}}{\partial k_s} = \frac{k_s}{E_N(k_s)}  ,   \label{eq:mubelow}    \\
\mu_N^{above} & = E_N(k_s)\, \frac{1 -  \frac{k_b^2E_N(k_b)}{k_s^2E_N(k_s)}\, \frac{\phi(k_s)}{\phi(k_b)}}{1 -  \frac{k_b^2}{k_s^2} \, \frac{\phi(k_s)}{\phi(k_b)}}  .   \label{eq:muabove}  
\end{align}
The ratio of $\mu_N^{above}$ to $\mu_N^{below}$ gives the gap of chemical potential at $\rho_{sat}$, and it is  
\begin{equation}
\frac{\mu_N^{above}}{\mu_N^{below}} = \frac{1 - \left[ \frac{r_1(k_s)E_N(k_b)}{r_1(k_b)E_N(k_s)} \right]  e^{- {\bar k}_s^2 + {\bar k}_b^2}}{1 - \left[ \frac{r_1(k_s)}{r_1(k_b)} \right] e^{- {\bar k}_s^2 + {\bar k}_b^2}} > 1 .   \label{eq:muratio}  
\end{equation}

We can calculate pressure, $P$, by using $\rho_N$, $\epsilon_N$, and $\mu_N$ 
\begin{equation}
P = \mu_N \rho_N - \epsilon_N  .  \label{eq:pressure}   
\end{equation}
Therefore, pressure has a gap at $\rho_{sat}$ as well. 

Finally, the (squared) sound velocity, $v_s^2$, is estimated by 
\begin{equation}
v_s^2 = \frac{\rho_N}{\mu_N} \left( \frac{\partial \mu_N}{\partial \rho_N} \right)  .  \label{eq:soundv}   
\end{equation}
Below $\rho_{sat}$, the sound velocity reads
\begin{equation}
v_s^{2\, below} = \frac{k_s^2}{3 E_N(k_s)^2}  .  \label{eq:belowvelocity}   
\end{equation}
To calculate $v_s^{2\, above}$, we need the second derivative, $\partial^2 k_b / \partial k_s^2$, which is again derived from the boundary condition, Eq.~(\ref{eq:determination1}).  We then obtain 
\begin{equation}
\frac{\partial^2 k_b}{\partial k_s^2} =  \frac{1}{\beta} \left[ \frac{\phi^\prime(k_s)}{\phi(k_b)} - \beta \left( \frac{\partial k_b}{\partial k_s} \right)^2 \frac{\phi^\prime(k_b)}{\phi(k_b)} \right] \xrightarrow{k_b \to 0} k_b^{-5}   ,   \label{eq:derkbks2}
\end{equation}
with
\begin{equation}
\phi^\prime(k) = \frac{N_c^3}{\sqrt{\pi}} 8{\bar k} [r_1(k) - r_2(k) {\bar k}^2] e^{- {\bar k}^2}  .  \label{eq:derphi}
\end{equation}
Thus, the sound velocity reads 
\begin{equation}
v_s^{2\, above} = \frac{\rho_N^{above}}{\mu_N^{above}} \left( \frac{\partial \mu_N^{above}}{\partial k_s} \right) \left/ \left( \frac{\partial \rho_N^{above}}{\partial k_s} \right) \right.  ,  \label{eq:abovevelocity}   
\end{equation}
where 
\begin{equation}
\frac{\partial \mu_N^{above}}{\partial k_s} = \mu_N^{above} \times \left( \frac{A^\prime}{A} - \frac{B^\prime}{B} \right) ,  \label{eq:ab1}   
\end{equation}
with 
\begin{align}
A & = k_s^2 E_N(k_s) - \beta k_b^2 E_N(k_b) \, \frac{\partial k_b}{\partial k_s}  ,  \label{eq:a1}    \\
B & = k_s^2 - \beta k_b^2 \, \frac{\partial k_b}{\partial k_s}  ,  \label{eq:b1}    \\
A^\prime & = 2 k_s E_N(k_s) + \frac{k_s^3}{E_N(k_s)} -\beta \left[ k_b^2 E_N(k_b)\, \frac{\partial^2 k_b}{\partial k_s^2} + \left( 2 k_b E_N(k_b) + \frac{k_b^3}{E_N(k_b)} \right) \left( \frac{\partial k_b}{\partial k_s} \right)^2 \, \right]  ,  \label{eq:a2}    \\
B^\prime & = 2 k_s -\beta \left[ k_b^2\, \frac{\partial^2 k_b}{\partial k_s^2} + 2 k_b \left( \frac{\partial k_b}{\partial k_s} \right)^2 \, \right] .  \label{eq:b2}      
\end{align}
%

\begin{figure}[t!]
  \includegraphics[width=15.0cm,keepaspectratio,clip]{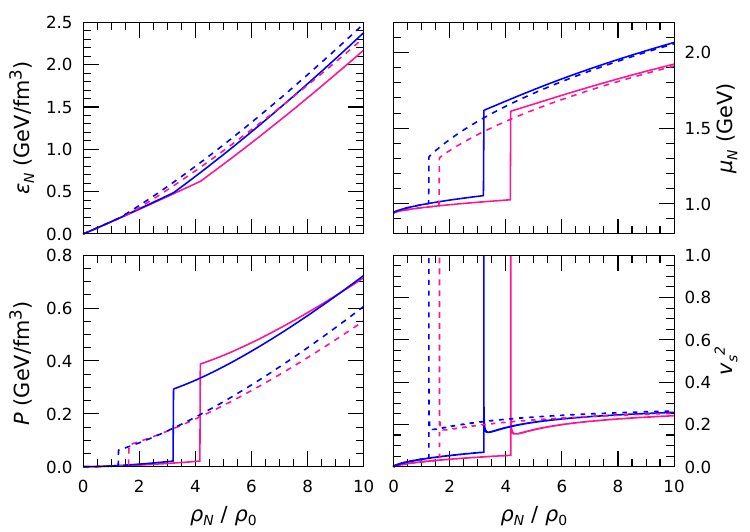}
  \caption{\label{fig:iniresult}  Energy density, chemical potential, pressure or sound velocity as a function of $\rho_N$.  The red solid (dashed) lines are for $r_p = 0.6 \, (0.8)$ fm in SNM, while the blue solid (dashed) lines are for $r_p = 0.6 \, (0.8)$ fm in PNM.  
  }
\end{figure}
In Fig.~\ref{fig:iniresult}, we show $\epsilon_N$, $\mu_N$, $P$ and $v_s^2$ in the case of $r_p = 0.6$ or $0.8$ fm.  At $\rho_{sat}$, the energy density is continuous, but is not smooth, while there is a large gap in the chemical potential or pressure, which may not be permitted in thermodynamics.  The sound velocity diverges at $\rho_{sat}$, as seen in the IdylliQ model.

\subsubsection{Pure neutron matter} \label{subsubsec:pnmresults1}

For PNM, the neutron density reads
\begin{align}
\rho_n^{below} & = \frac{1}{3\pi^2} k_s^3  \ \ \ \ \ \ \  (k_s = k_F),   \label{eq:rhobelown}    \\
\rho_n^{above} & = \frac{1}{3\pi^2}  ( k_s^3 - \beta^{\prime} k_b^3 )   ,   \label{eq:rhoaboven}  
\end{align}
and the energy density is 
\begin{align}
\epsilon_n^{below} & = \frac{1}{\pi^2} \int_0^{k_s} dk \, k^2 E_n(k) = \frac{1}{\pi^2}  I(M_N, k_s) ,  \label{eq:energybelown}    \\
\epsilon_n^{above} & = \frac{1}{\pi^2}  \left( \int_0^{k_s} - \beta^{\prime} \int_0^{k_b} \right) dk \, k^2 E_n(k) = \frac{1}{\pi^2}  [ I(M_N, k_s) - \beta^{\prime} I(M_N, k_b) ] ,    \label{eq:energyaboven}  
\end{align}
with $E_n(k) = \sqrt{M_N^2 + k^2}$. 
Then, replacing $N_c^3$ with $\frac{2}{3} N_c^3$, we can calculate $\mu_n$, $P$, and $v_s^2$ as in case of SNM.   Fig.~\ref{fig:iniresult} depicts $\epsilon_n$, $\mu_n$, $P$, and $v_s^2$ in PNM as well.

\subsection{Gaussian quarkyonic model with minimal corrections} \label{subsec:correctedgaussianmodel}

As seen in Section\,\ref{subsec:results1}, we have encountered the singular behavior of physical quantities in the na{\"i}ve GQ model.  We thus need to somehow remedy such behavior.  Because the cause of gap and divergence at the quark saturation density comes from $\phi(k_b)$ in Eq.~(\ref{eq:derkbks}), which is related to $\partial g_{Q\, (d)}(k_b)/\partial k_b$, a minimal correction to cure this problem can be provided with a smearing of the sharp Fermi surface, $\theta(k_b - k)$, in the nucleon distributions, Eqs.~(\ref{eq:nucleondist1}) and (\ref{eq:nucleondist2}).  Such a smearing should be originally evaluated by the quark and gluon exchanges among nucleons, but it is extremely difficult to handle them in a multi-baryon system.  Instead, for example, we could apply Sommerfeld expansion to blur the Fermi surface phenomenologically -- for details, see Appendix\,\ref{app:sommerfeld}.  

\begin{figure*}[h!]
 \includegraphics[width=12.0cm,keepaspectratio,clip]{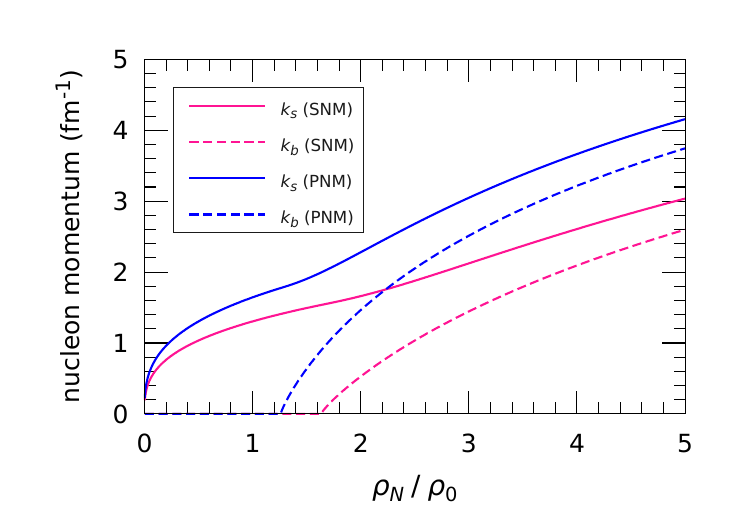}%
  \caption{\label{fig:correctkbks}
   Density dependence of $k_s$ and $k_b$ in the case of $r_p=0.8$ fm, $\nu =1.8$ and $w=0.25$ GeV. 
   }
\end{figure*}

In this paper, we however take another, more practical prescription to remove the singular behavior in the present model.  Here, we want to modify $k_b^2$ in the denominator of Eq.~(\ref{eq:derkbks}) so that the power of $k_b$ is weaker.  In the method of Sommerfeld expansion, the power of $k_b$ is eventually weakened (see Appendix\,\ref{app:sommerfeld}).  To do so, we shall introduce an infrared regulator to the boundary conditions, Eqs.~(\ref{eq:determination1}) and (\ref{eq:determination2}).  Then, we take the simplest choice  
\begin{equation}
\alpha_\nu(k_b) = 1 + \left( \frac{w}{k_b} \right)^\nu  \to \infty \, (1) \ {\rm as} \ k_b \to 0 \,  (\infty),  \label{eq:regulator1}   
\end{equation}
with $w$ a width parameter and $\nu$ a positive, real number.  The boundary conditions are thus redefined by 
\begin{align}
g_Q(k_s) &- \beta \alpha_\nu(k_b) g_Q(k_b) = 1   \ \ \ \ \  {\rm for \ SNM},   \label{eq:determination3}    \\
g_d(k_s) &- \beta^{\prime} \alpha_\nu(k_b) g_d(k_b) = 1   \ \ \ \ \ {\rm for \ PNM}  .  \label{eq:determination4}  
\end{align}
Note that, when $w=0$,  $\alpha_{\nu} = 1$ (without the regulator).  
Because $g_{Q\, (d)}(k_b)$ behaves like ${\cal O}(k_b^3)$ when $k_b \to 0$, the parameter, $\nu$, should be taken as $0 < \nu < 3$.  Furthermore, we find 
\begin{align}
\alpha_\nu^\prime(k_b) &\equiv \frac{d}{d{\bar k}_b} \alpha_\nu(k_b) = - \frac{\nu}{\bar w}  \left( \frac{w}{k_b} \right)^{\nu +1} ,   \label{eq:derregulator1}    \\
\alpha_\nu^{\prime \prime}(k_b) &\equiv \frac{d^2}{d{\bar k}_b^2} \alpha_\nu(k_b) = \frac{\nu (\nu +1)}{{\bar w}^2}  \left( \frac{w}{k_b} \right)^{\nu +2}  ,   \label{eq:derregulator2}  
\end{align}
where ${\bar w} (\equiv aw/N_c)$, $\alpha_\nu$, $\alpha_\nu^\prime$, and $\alpha_\nu^{\prime \prime}$ are all dimensionless.  Fig.~\ref{fig:correctkbks} depicts the momenta, $k_b$ and $k_s$, calculated with the modified boundary condition, Eqs.~(\ref{eq:determination3}) and (\ref{eq:determination4}).  We notice that the momenta increase gradually beyond the quark saturation density, compared with the result without the regulator.  

In the case of SNM, using the modified boundary condition, the first derivative, $\partial k_b / \partial k_s$, reads  
\begin{equation}
\frac{\partial {\bar k}_b}{\partial {\bar k}_s} = \frac{\partial k_b}{\partial k_s} = \frac{\phi(k_s)}{\beta [\alpha_\nu^\prime(k_b) g_Q(k_b) + \alpha_\nu(k_b) \phi(k_b)]}  \xrightarrow{k_b \to 0} k_b^{\nu -2}  ,   \label{eq:modderkbks1}
\end{equation}
which ensures that the physical quantities (except $v_s^2$) are continuous at the quark saturation density.  Similarly, using the modified boundary condition again, the second derivative reads
\begin{equation}
\frac{\partial^2 {\bar k}_b}{\partial {\bar k}_s^2} = \left( \frac{N_c}{a} \right) \frac{\partial^2 k_b}{\partial k_s^2} = \frac{\phi^\prime(k_s) -\beta \left( \frac{\partial {\bar k}_b}{\partial {\bar k}_s} \right)^2 X(k_b)}{\beta [\alpha_\nu^\prime(k_b) g_Q(k_b) + \alpha_\nu(k_b) \phi(k_b)]}  \xrightarrow{k_b \to 0} k_b^{2\nu -5}  ,   \label{eq:modder2kbks1}
\end{equation}
with 
\begin{equation}
X(k_b) \equiv \frac{\partial^2}{\partial {\bar k}_b^2} \alpha_\nu(k_b) g_Q(k_b) = \alpha_\nu^{\prime \prime}(k_b) g_Q(k_b) + 2 \alpha_\nu^\prime(k_b) \phi(k_b) + \alpha_\nu(k_b) \phi^\prime(k_b)  .     \label{eq:x1}
\end{equation}
Therefore, to make the sound velocity continuous and make $\mu_N$ and $P$ smooth at $\rho_{sat}$, the parameter, $\nu$, is further restricted to be $\frac{3}{2} < \nu < 3$ (see Eqs.~(\ref{eq:a2}) and (\ref{eq:b2})).  For PNM, it is necessary to replace $\beta$ with $\beta^\prime$.  

\begin{figure*}[h!]
 \includegraphics[width=15.0cm,keepaspectratio,clip]{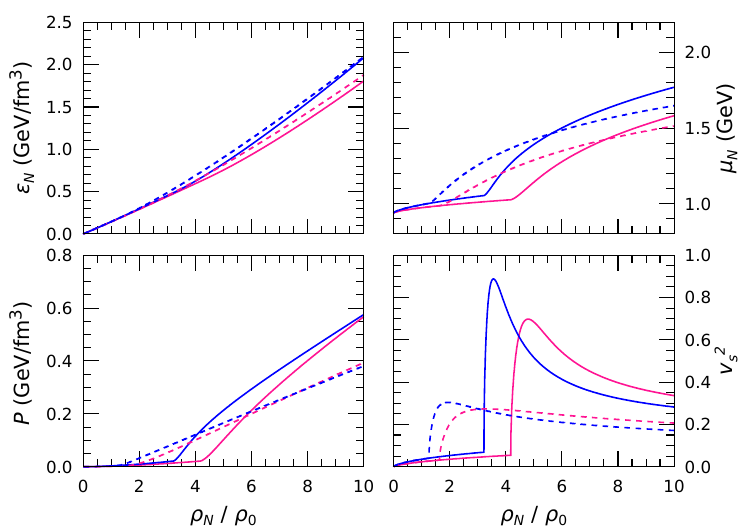}%
  \caption{\label{fig:modsnmresult}
  Energy density, chemical potential, pressure and sound velocity in the case of $\nu =1.8$ and $w=0.25$ GeV.  The red solid (dashed) lines are for $r_p = 0.6 \, (0.8)$ fm in SNM, while the blue solid (dashed) lines are for $r_p = 0.6 \, (0.8)$ fm in PNM.  
   }
\end{figure*}
Now, we can calculate $\epsilon_N$, $\mu_N$, $P$, and $v_s^2$, as in Section\,\ref{subsec:results1}, replacing $\partial k_b / \partial k_s$ and $\partial^2 k_b/\partial k_s^2$ with the modified ones.  
In Fig.~\ref{fig:modsnmresult}, we present the result in SNM or PNM.  As expected, $\epsilon_N$, $\mu_N$, $P$, and $v_s^2$ are all continuous, and $\epsilon_N$, $\mu_N$ and $P$ are smooth at $\rho_{sat}$.   

\begin{figure*}[h!]
 \includegraphics[width=15.0cm,keepaspectratio,clip]{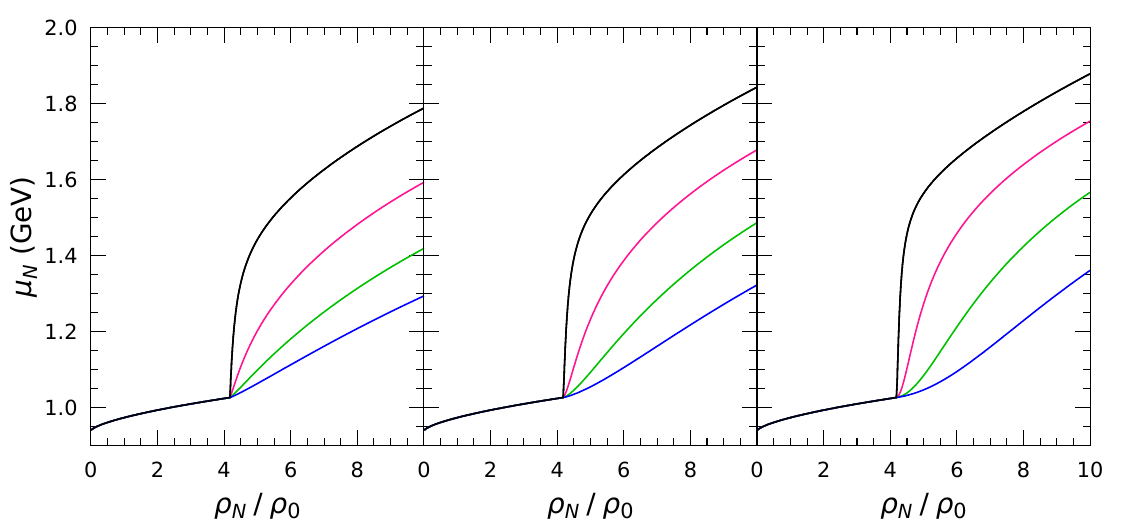} \\
  \includegraphics[width=15.0cm,keepaspectratio,clip]{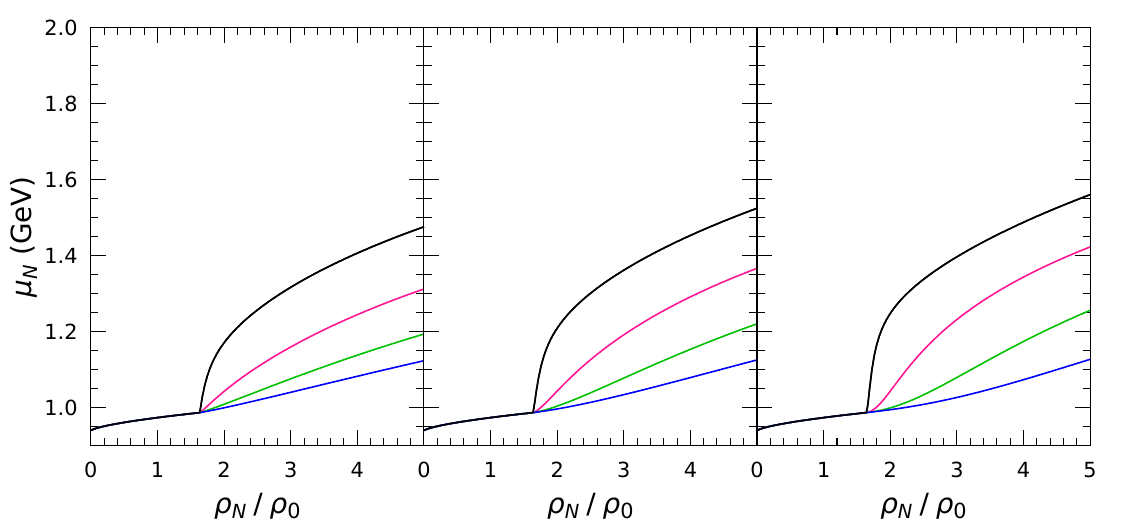}
  \caption{\label{fig:modmu}
 Dependence of chemical potential on $\nu$ and $w$ in SNM.  The upper (lower) panels are for $r_p = 0.6 \, (0.8)$ fm.  The left (middle) [right] panels are for $\nu = 1.6 \, (1.8) \, [2.0]$.  The black (red) [green] \{blue\} lines are for $w = 0.1 \, (0.2) \, [0.3] \, \{0.4 \}$ GeV.  
   }
\end{figure*}
\begin{figure*}[h!]
 \includegraphics[width=15.0cm,keepaspectratio,clip]{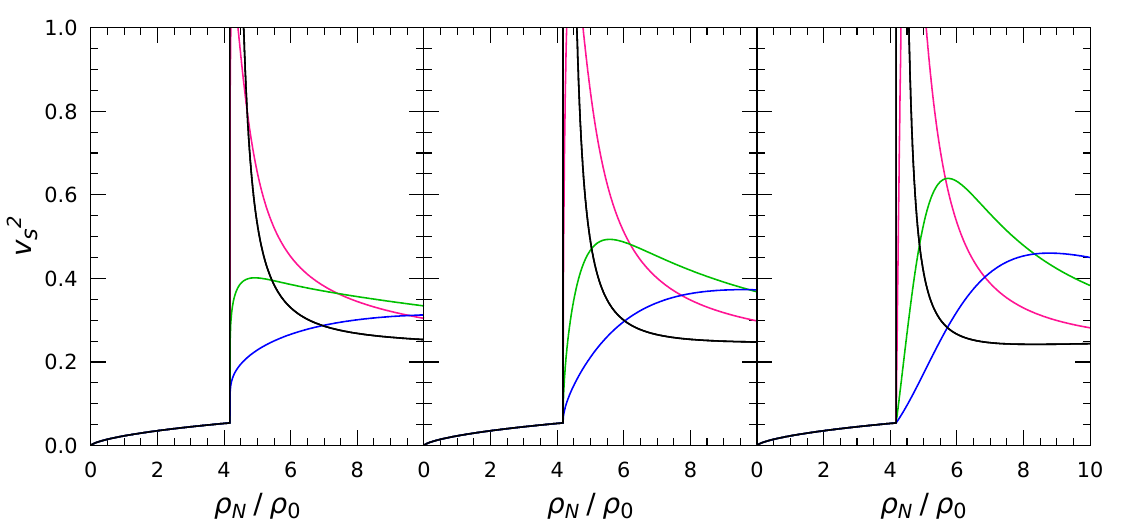} \\
 \includegraphics[width=15.0cm,keepaspectratio,clip]{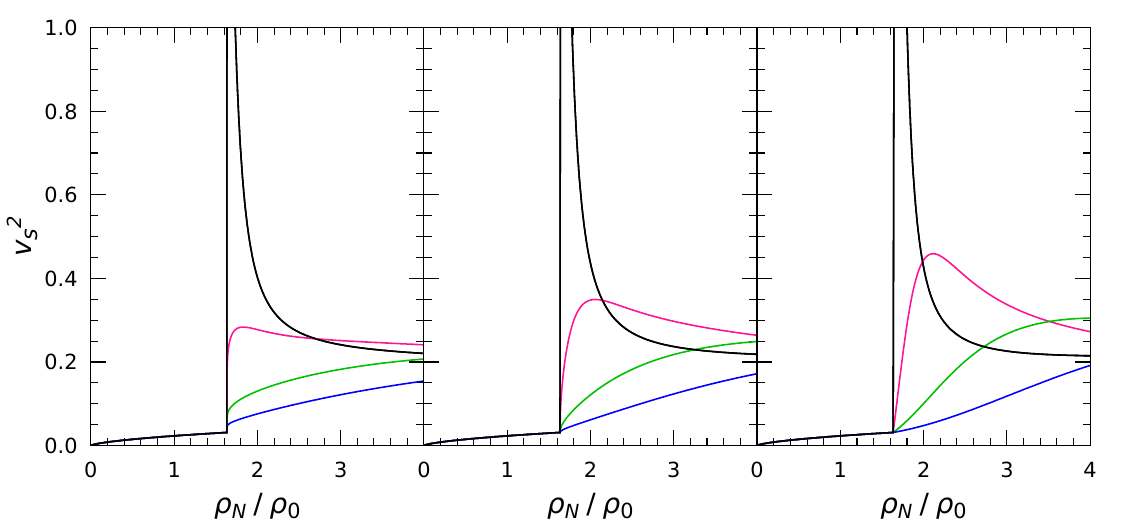}
  \caption{\label{fig:modsound}
  Dependence of sound velocity on $\nu$ and $w$ in SNM.  The upper (lower) panels are for $r_p = 0.6 \, (0.8)$ fm.  The left (middle) [right] panels are for $\nu = 1.6 \, (1.8) \, [2.0]$.  The black (red) [green] \{blue\} lines are for $w = 0.1 \, (0.2) \, [0.3] \, \{0.4 \}$ GeV. 
   }
\end{figure*}
Furthermore, Fig.~\ref{fig:modmu} and \ref{fig:modsound} respectively show the dependence of $\mu_N$ and $v_s^2$ on $\nu$ and $w$ in detail.  It is seen that, in the chemical potential, as the width parameter, $w$, is larger, $\mu_N$ beyond $\rho_{sat}$ is more suppressed, and that the larger power, $\nu$, tames the rapid rise at $\rho_{sat}$ more.  We can see the similar trend in $v_s^2$ as well.  The sound velocity with $w=0.1$ GeV does not diverge at $\rho_{sat}$, but overshoots the upper limit.  The same occurs in the case of $r_p=0.6$ fm and $w=0.2$ GeV.  The parameter, $w$, broadens the peak of the sound velocity, while $\nu$ enhances the height of the peak and hinders the rapid increase of the velocity around $\rho_{sat}$.

\clearpage

\newpage
\section{Quarkyonic quark-meson coupling model} \label{sec:qqmc}

We here build a new, quark-based nuclear model, in which we incorporate the GQ model into the QMC model.  The two models are combined each other through the gaussian quark wavefunction, Eq.~(\ref{eq:sol}).  We will call it the quarkyonic quark-meson coupling (QQMC) model.  

\subsection{Brief review of the QMC model} \label{subsec:revqmc}

In this section, we briefly review the quark-meson coupling (QMC) model with the gaussian quark wavefunction.  

We consider the mean fields of $\sigma$, $\omega$, and $\rho$ mesons, which interact with the confined quarks, in uniformly distributed, asymmetric nuclear matter. 
Let the mean-field values for the $\sigma$, $\omega$ (the time component) and $\rho$ (the time component in the 3rd direction of isospin) fields be ${\bar \sigma}$, ${\bar \omega}$ and ${\bar \rho}$, respectively.  

Under the relativistic HO potential, Eq.~(\ref{eq:conpot}), the Dirac equation for the quark field $\psi_j$ ($j= u\, \textrm{or}\, d$) is given by~\cite{PhysRevD.110.113001}
\begin{equation}
  \left[i\gamma \cdot \partial -\left(m -V_{s}\right)-\gamma_{0}V_{0}-\frac{c}{2}\left( 1+\gamma_{0}\right) r^2 \right]\psi_{j}({\bm r}, t)=0 , 
  \label{eq:Dirac-eq1}
\end{equation}
where $V_{s} = g^{q}_{\sigma}\bar{\sigma}$ and $V_{0} = g^{q}_{\omega}\bar{\omega} + \tau_{3j} g^{q}_{\rho} \bar{\rho}$ [$\tau_{3j} = \pm 1$ for $\binom{u}{d}$ quark] with the quark-meson coupling constants, $g^{q}_{\sigma}$, $g^{q}_{\omega}$ and $g^{q}_{\rho}$.  Note that the isoscalar $\sigma$ meson couples to the $u$ and $d$ quarks equally.  We respectively define the effective quark mass and the effective single-particle quark energy as $m^{\ast} \equiv m-V_{s} = m - g^{q}_{\sigma}\bar{\sigma}$ and $\epsilon^{\ast} \equiv \epsilon_j -V_{0} = \epsilon_j - g^{q}_{\omega}\bar{\omega} \mp g^{q}_{\rho} \bar{\rho}$ for $\binom{u}{d}$ quark, where $\epsilon_{j}$ is the eigenenergy of Eq.~(\ref{eq:Dirac-eq1}).  The static, lowest-state wavefunction in matter is presented by
\begin{equation}
  \psi_{j}({\bm r}, t)
  = \exp\left[ -i\epsilon_{j}t \right] \psi_Q({\bm r})  .
  \label{eq:sol1}
\end{equation}
The wavefunction $\psi_Q({\bm r})$ is then given by Eqs.~(\ref{eq:sol}) and (\ref{eq:norm}), in which $\epsilon$, $m$, $\lambda$ and $a$ are, respectively, replaced with $\epsilon^{\ast}$, $m^{\ast}$, $\lambda^{\ast}$ and $a^{\ast}$, and the effective energy $\epsilon^{\ast}$ is determined by $\sqrt{\epsilon^\ast + m\ast} (\epsilon^\ast - m^\ast) = 3 \sqrt{c}$ (see Eq.~(\ref{eq:qenergy})).  We assume that the strength parameter, $c$, of the confinement potential does not change in matter.  

The zeroth-order energy of the nucleon in matter is thus given by $E_N^{0 \ast} = 3 \epsilon^\ast$, and we have the effective nucleon mass in matter ($M_N^\ast = M_p^\ast = M_n^\ast$)
\begin{equation}
  M_N^\ast = E_B^{0 \ast} + E^{spin}_N - E^{c.m. \ast}_N . 
  \label{eq:nmass1}
\end{equation}
Here, we assume that the spin correlation $E^{spin}_N$ does not change in matter, and the c.m. correction $E^{c.m. \ast}_N$ is given by Eq.~(\ref{eq:cmcorrection}) with $\epsilon^{\ast}$ and $m^{\ast}$, instead of $\epsilon$ and $m$.  

For describing asymmetric nuclear matter, we now start from the following Lagrangian density in mean-field approximation~\cite{GUICHON1996349}
\begin{align}
  {\cal L} = &{\bar \psi}_N \left[ i \gamma \cdot \partial - M^\ast_N({\bar \sigma}) - g_\omega \gamma_0 {\bar \omega} - g_\rho \gamma_0 \tau_3 {\bar \rho} \right] \psi_N  \nonumber \\
  & - \frac{m_\sigma^2}{2} {\bar \sigma}^2 + \frac{m_\omega^2}{2} {\bar \omega}^2 + \frac{m_\rho^2}{2} {\bar \rho}^2 - \frac{g_2}{3} {\bar \sigma}^3 
  \label{eq:lagrangian}
\end{align}
with $\psi_N$ the nucleon field, and $\tau_3$ the 3rd component of Pauli matrix.  The nucleon-meson coupling constants, $g_\sigma$,  $g_\omega$, and $g_\rho$, are respectively related to the quark-meson coupling constants as $g_\sigma = 3g_\sigma^q$,  $g_\omega = 3g_\omega^q$, and $g_\rho = g_\rho^q$.  The meson masses are taken to be $m_\sigma = 550$ MeV, $m_\omega = 783$ MeV, and $m_\rho = 770$ MeV.  We add the last term to the Lagrangian, which is the nonlinear, self-coupling term of $\sigma$ meson, in order to reproduce the properties of nuclear matter as discussed later. Here, we do not include the nonlinear term $\frac{1}{4} g_3 \sigma^4$, because it plays similar roles as $\frac{1}{3} g_2 \sigma^3$.  

The total energy per nucleon of asymmetric nuclear matter is then obtained by 
\begin{align}
  \epsilon_{tot}/\rho_N
  = \ &\epsilon_N^\ast/\rho_N + g_{\omega}\bar{\omega} + g_{\rho} \left( \frac{\rho_3}{\rho_N} \right) \bar{\rho} \nonumber \\
  &+ \frac{1}{2\rho_N}\left(m_{\sigma}^{2}\bar{\sigma}^{2}-m_{\omega}^{2}\bar{\omega}^{2}-m_{\rho}^{2}\bar{\rho}^{2}\right)
    +  \frac{1}{3\rho_N}g_{2}\bar{\sigma}^{3} , 
    \label{eq:energydensity}
\end{align}
where 
\begin{equation}
\epsilon_N^\ast = \frac{1}{\pi^{2}} \int dk \,k^{2} E_N^\ast(k) 
\left[ f_p(k) + f_n(k) \right] ,  \label{eq:energykin}
\end{equation}
with $E_N^\ast(k) = \sqrt{M_N^{\ast 2}+k^{2}}$.  Note that $f_{i=p,\, n}(k) = \theta(k_{F_i} -k)$ with $k_{F_i}$ the Fermi momentum for protons or neutrons, which is related to the density of protons (neutrons), $\rho_{p (n)}$, through $\rho_{p (n)} = k_{F_{p (n)}}^3/(3\pi^2)$. 
The total nucleon density is given by $\rho_N = \rho_p + \rho_n$, and the difference in proton and neutron densities is defined by $\rho_3 \equiv \rho_p - \rho_n$.  The binding energy per nucleon, $\epsilon_b$, is given by $\epsilon_{b}(\rho_{N}) = \epsilon_{tot}/\rho_N - M_N$.  

The mean-field values of the meson fields satisfy~\cite{PhysRevD.110.113001} 
\begin{align}
  \left(m_{\sigma}^{2}+g_{2}\bar{\sigma} \right) \bar{\sigma}
  &= - \left( \frac{\partial M_N^\ast}{\partial {\bar \sigma}} \right) ( \rho_{p}^{s} + \rho_{n}^{s} )
  \equiv g_\sigma G_{\sigma}(\bar{\sigma}) ( \rho_{p}^{s} + \rho_{n}^{s} ) , 
  \label{eq:sigmamf} \\
  m_{\omega}^{2} \bar{\omega} &= g_{\omega}\rho_{N},  \label{eq:omegamf} \\
  m_{\rho}^{2}\bar{\rho} &= g_{\rho} \rho_{3}  ,  \label{eq:rhomf}
\end{align}
where Eq.~(\ref{eq:sigmamf}) is the gap equation for the nucleon mass in matter.   
Here, $\rho_{i=p (n)}^{s}$ is the scalar density of protons (neutrons) in matter 
\begin{equation}
 \rho_{i}^{s} = \frac{1}{\pi^{2}} \int dk \,k^{2} \frac{M_{N}^{\ast}}{\sqrt{M_{N}^{\ast 2}+k^{2}}} f_i(k) = \frac{M_{N}^{\ast}}{\pi^{2}} J(M_{N}^{\ast}, k_{F_i}) . 
  \label{eq:scalardenj}
\end{equation}
with
\begin{align}
J(M, x) &= \int_0^x dt \, \frac{t^2}{\sqrt{M^2 + t^2}}    \nonumber \\
&= \frac{1}{2} \left[ x \sqrt{M^2 + x^2} + M^2 \log \left( \frac{M}{x + \sqrt{M^2 + x^2}} \right) \right] .   \label{eq:integ2}
\end{align}
Furthermore, $G_{\sigma}({\bar \sigma})$ is {\it the isoscalar, Lorentz scalar polarizability}~\cite{PhysRevD.110.113001} 
\begin{equation}
G_{\sigma}({\bar \sigma}) \equiv S^\ast({\bar \sigma}) - \frac{1}{3} \frac{\partial}{\partial m^\ast} E_N^{c.m. \ast}  ,    \label{eq:polar1}
\end{equation}
where  
\begin{align}
S^\ast({\bar \sigma}) &= 
\int_N d{\bm r} \, {\bar \psi}_Q({\bm r}) \psi_Q({\bm r}) = 
\frac{\partial \epsilon^\ast}{\partial m^\ast} = \frac{\epsilon^\ast + 3m^\ast}{3\epsilon^\ast + m^\ast}    ,     \label{eq:polar2}   \\
\frac{1}{3} \frac{\partial}{\partial m^\ast} E_N^{c.m. \ast} & = \frac{1}{27} \left[ 
\frac{2 (\epsilon^\ast S^\ast - m^\ast)(77\epsilon^\ast + 31m^\ast)}{(3\epsilon^\ast + m^\ast)^2} 
+ \frac{(\epsilon^{\ast\, 2} - m^{\ast\, 2})(77 S^\ast + 31)}{(3\epsilon^\ast + m^\ast)^2}  \right.  \nonumber \\
&\left. -2 \frac{(\epsilon^{\ast\, 2} - m^{\ast\, 2})(77 \epsilon^\ast + 31m^\ast)(3 S^\ast +1)}{(3\epsilon^\ast + m^\ast)^3}   \right]    .    \label{eq:polar3}
\end{align}

Now, we have four parameters, $g_\sigma$, $g_\omega$, $g_\rho$ and $g_2$.  
The coupling constants, $g_\sigma$ and $g_\omega$, are determined so as to fit the binding energy, $\epsilon_b(\rho_0) = -16$ MeV, at the nuclear saturation density in  equilibrium, symmetric nuclear matter.  
We fit the $\rho$-nucleon coupling constant so as to reproduce the bulk symmetry energy of symmetric nuclear matter, $E_{sym}(\rho_0) = 32.0$ MeV.  Finally, the nonlinear coupling constant, $g_2$, is used to fit the nuclear incompressibility, $K_0 = 240$ MeV.  

\begin{table}[t!]
  \caption{\label{tab:QMCcc}
   Coupling constants in the case 2.  For the detail, see the text. 
  }
  \begin{ruledtabular}
    \begin{tabular}{ccccc}
      $r_p$ (fm) & $g_{\sigma}$ & $g_{\omega}$ & $g_{\rho}$ & $g_{2}$ (fm$^{-1}$) \\
      \colrule
      0.6                             & 11.00        & 7.47         & 4.36       & 22.78               \\
      0.7                             & 10.49        & 7.56         & 4.36       & 23.81               \\
      0.8                             & 10.11        & 7.65         & 4.35       & 24.65               \\
      0.9                             &  9.83        & 7.74         & 4.35       & 25.32               \\
    \end{tabular}
  \end{ruledtabular}
\end{table}

\begin{table}[t!]
  \caption{\label{tab:properties}
  Properties of in-medium nucleons and symmetric nuclear matter at the nuclear saturation density, $\rho_0$, in the case 2.
    Here, $L$ and $K_{sym}$ are, respectively, the slope and the curvature parameters of nuclear symmetry energy.  The quark saturation densities in matter are listed in the last two columns (see Section\,\ref{subsec:qqmc}).  
  }
  \begin{ruledtabular}
    \begin{tabular}{ccccccccccc}
      $r_p$ & $r_p^{\ast}$ & $m^{\ast}$ & $\epsilon^{\ast}$ & $M_{N}^{\ast}$ & $E_{N}^{0\ast}$ & $E_{N}^{c.m.\ast}$ & $L$  & $K_{\textrm{sym}}$ & \multicolumn{2}{c}{$\rho_{\textrm{sat}}^\ast/\rho_0$} \\
      \cline{1-2}\cline{3-9}\cline{10-11}
      \multicolumn{2}{c}{(fm)}                                    & \multicolumn{7}{c}{(MeV)}                                                                                 &  SNM  &   PNM        \\
      \colrule
      0.6                        & 0.627                          & 223.5      & 620.4             & 776.5          & 1861.2          & 468.4              & 86.4 & $-18.1$            & 3.594                                & 2.826                                \\
      0.7                        & 0.733                          & 226.7      & 541.6             & 773.8          & 1624.9          & 382.2              & 86.5 & $-17.5$            & 2.212                                & 1.740                                \\
      0.8                        & 0.840                          & 228.9      & 484.7             & 771.1          & 1454.1          & 318.0              & 86.6 & $-16.9$            & 1.459                                & 1.146                                \\
      0.9                        & 0.946                          & 230.5      & 442.1             & 768.6          & 1326.4          & 268.7              & 86.7 & $-16.3$            & 1.012                                & 0.794                                \\
    \end{tabular}
  \end{ruledtabular}
\end{table}

The coupling constants and the properties of nucleons and nuclear matter in the case 2 are, respectively, listed in Tables~\ref{tab:QMCcc} and \ref{tab:properties}.  At $\rho_0$, the root-mean-square charge radius of proton, $r_p^\ast$, increases by about $5 \%$ compared with the radius in vacuum, $r_p$.  This swelling is consistent with the experimental data (see Ref.~\cite{LU1998217}).  We also calculate the slope and the curvature parameters of nuclear symmetry energy, $L$ and $K_{sym}$, which are related to astrophysical multi-messenger observations of neutron stars~\cite{Miyatsu_2022, MIYATSU2023138013}.  

\begin{figure}[t!]
  \includegraphics[width=8.0cm,keepaspectratio,clip]{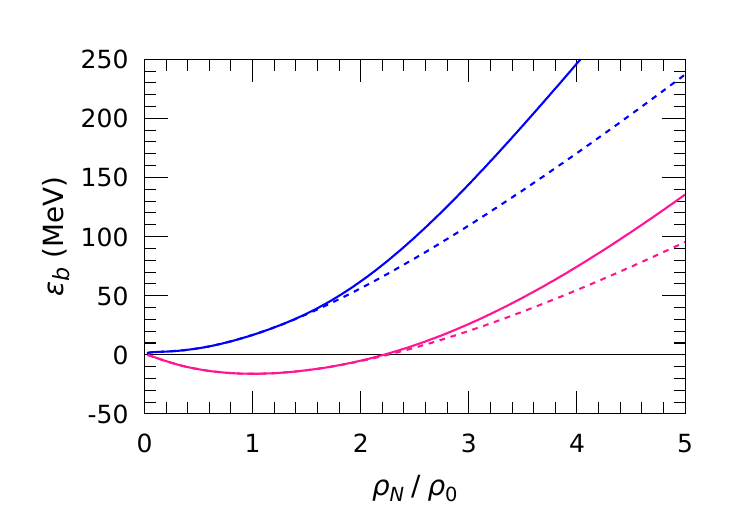}
  \includegraphics[width=8.0cm,keepaspectratio,clip]{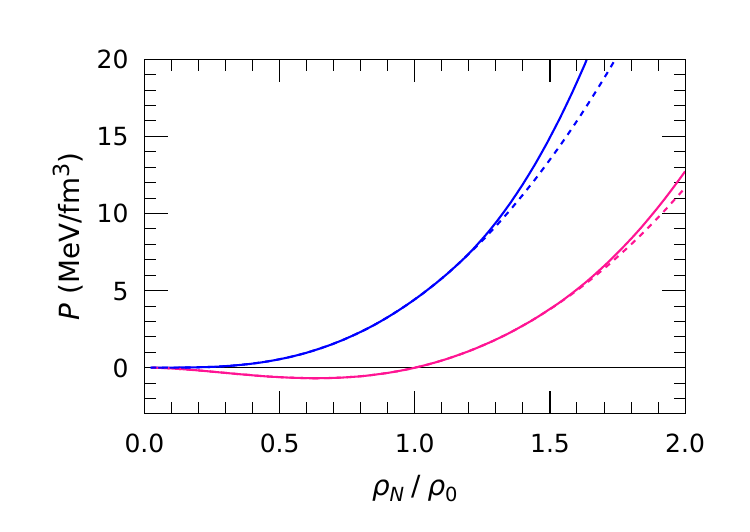}
  \caption{\label{fig:Eb-Press}  Binding energy per nucleon, $\epsilon_{b}$, and pressure, $P$, in the case of $r_p=0.8$ fm, $\nu = 2.0$ and $w=0.4$ GeV.  The red solid (dashed) lines denote the results of the QQMC (QMC) model in SNM, while the blue solid (dashed) lines denote those of the QQMC (QMC) model in PNM. 
  }
\end{figure}
\begin{figure}[h!]
 \includegraphics[width=12.0cm,keepaspectratio,clip]{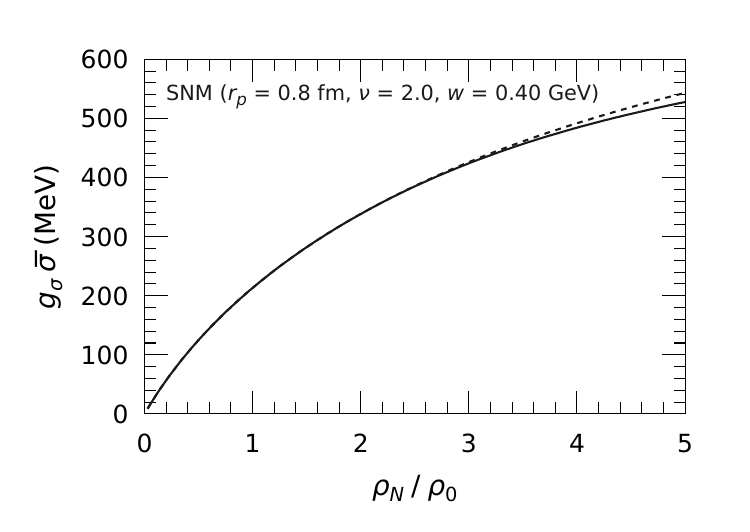}
  \caption{\label{fig:sigma-fields}
   Density dependence of the sigma field, $g_\sigma \bar{\sigma}$, in the case of $r_p=0.8$ fm.  The solid (dashed) line shows the result of the QQMC (QMC) model. 
   }
\end{figure}
\begin{figure}[h!]
 \includegraphics[width=12.0cm,keepaspectratio,clip]{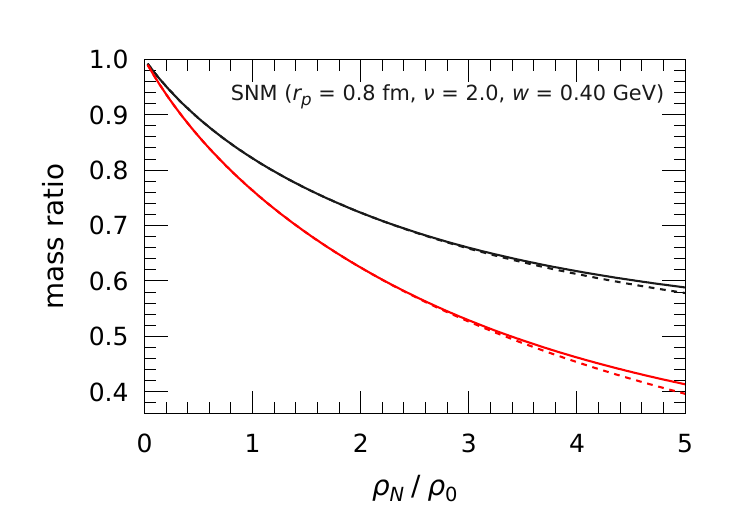}
  \caption{\label{fig:Quark-Nucleon-mass}
   Density dependence of mass ratio, $M_N^\ast / M_N$ (black lines) or $m^\ast / m$ (red lines), in the case of $r_p=0.8$ fm.  The solid (dashed) lines show the results of the QQMC (QMC) model.  
   }
\end{figure}
\begin{figure}[h!]
 \includegraphics[width=12.0cm,keepaspectratio,clip]{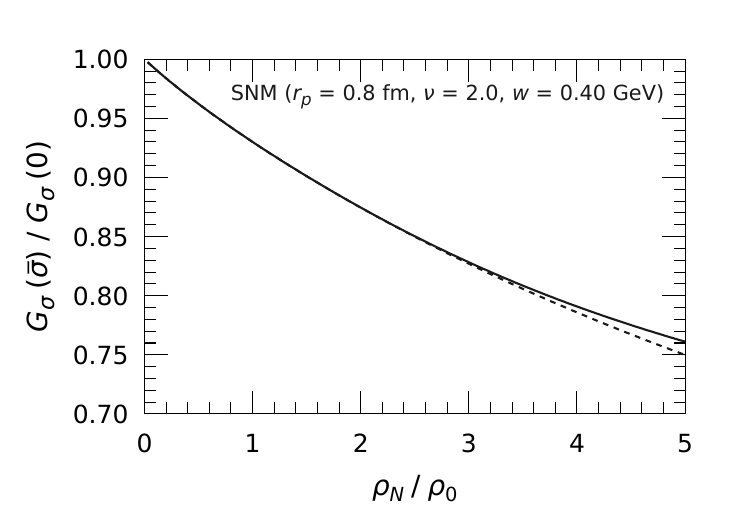}
  \caption{\label{fig:Gsigma}
   Density dependence of the scalar polarizability, $G_{\sigma}(\bar{\sigma}) / G_{\sigma}(0)$, in the case of $r_p=0.8$ fm.  The solid (dashed) line denotes the result of the QQMC (QMC) model.  
   }
\end{figure}

The binding energy per nucleon and pressure are showed in Fig.~\ref{fig:Eb-Press}.  Together with the results of the QMC model, those of the QQMC model (see Sec.~\ref{subsec:qqmc}) are illustrated in the figure as well.  Fig.~\ref{fig:sigma-fields} illustrates the mean-field value for the $\sigma$ meson in SNM.  The increase of the scalar field becomes slower at high density, compared with that at low density, which generates the nuclear saturation with a moderate value of incompressibility, $K_0$~\cite{GUICHON1988235}.  In Fig.~\ref{fig:Quark-Nucleon-mass}, we present the density dependence of the effective quark mass or that of the effective nucleon mass in SNM.  
Fig.~\ref{fig:Gsigma} depicts the scalar polarizability, $G_{\sigma }$.  In matter, the  quark mass is lighter, i.e.  {\it more relativistic}, than the mass in vacuum.  Therefore, the lower component in the quark wavefunction is enhanced, and the quark scalar density in a nucleon in matter is more reduced, i.e. $G_{\sigma}(0) > G_{\sigma}(\bar{\sigma})$, which softens the incomplessibility, $K_0$.   

\clearpage

\subsection{Quarkyonic QMC model} \label{subsec:qqmc}

In this Section, we construct the QQMC model for symmetric nuclear matter and pure neutron matter, in which the nuclear interaction is involved at the level of mean-field approximation.  
Furthermore, in the QMC model, as in Quantum Hadrodynamics (QHD), the physical quantities like the nuclear density, the energy density, etc. in matter can be expressed by those of a relativistic Fermi gas of nucleons of mass $M^\ast$ and both quark and nucleon degrees of freedom couple to the same mean scalar and vector fields.  This implies that duality remains valid below the onset of hard deconfinement.  Thus, we assume duality and formulate the QQMC model in terms of the nucleon degrees of freedom only.   

First, the quark saturation density, $\rho_{sat}^\ast$, in SNM is determined by Eq.~(\ref{eq:gQ0}) with $m^\ast$, $\epsilon^\ast$, and $a^\ast$, instead of $m$, $\epsilon$, and $a$: 
\begin{equation}
g_Q^\ast(k_{sat}^\ast) = \frac{N_c^3}{\sqrt{\pi}} \left[ \sqrt{\pi} - 2{\rm Er}({\bar k}_{sat}^\ast) 
- 2 {\bar k}_{sat}^\ast \left( 1 + \frac{2{\bar k}_{sat}^{\ast\, 2}}{2{\bar \lambda}^{\ast\, 2}+3} \right) e^{-{\bar k}_{sat}^{\ast\, 2}} \right] = 1   ,  \label{eq:gQ1}
\end{equation}
with ${\bar \lambda}^\ast = a^\ast \lambda^\ast$ and ${\bar k}^\ast = a^\ast k^\ast/N_c$ dimensionless variables.  Then, we find $\rho_{sat}^\ast = 2k_{sat}^{\ast\, 3}/(3\pi^2)$, which is smaller than $\rho_{sat}$ in the GQ model, because the quark wavefunction in matter is more spread out than in free space, i.e. $a^\ast > a$.  For PNM, Eq.~(\ref{eq:gd0}) with $m^\ast$, $\epsilon^\ast$, and $a^\ast$ provides the quark saturation density.  In the last two columns of Table~\ref{tab:properties}, the quark saturation density is summarized.  In fact, $\rho_{sat}^\ast$ is roughly $10\%$ smaller than $\rho_{sat}$.  

Once the quark saturation density is found, we can realize the model in the pre and  postsaturation regions as in Sec.~\ref{subsec:correctedgaussianmodel}.  
In the presaturation region, i.e. $\rho_N < \rho_{sat}^\ast$, the matter can be well described by the QMC model itself.  

\begin{figure*}[h!]
 \includegraphics[width=12.0cm,keepaspectratio,clip]{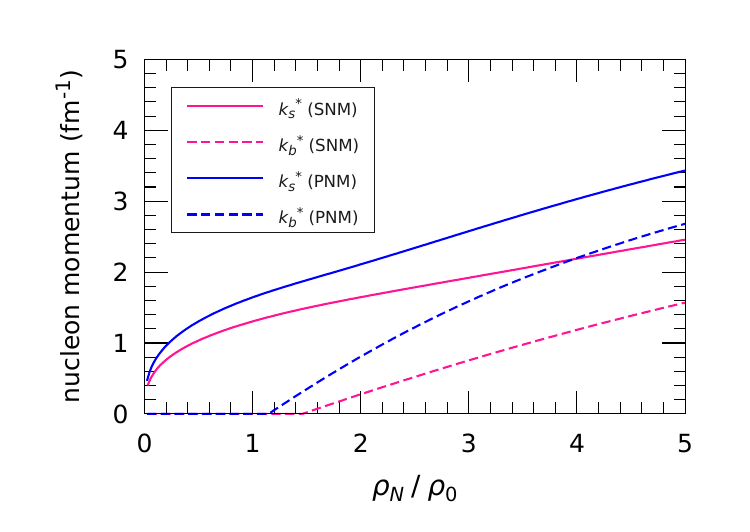}%
  \caption{\label{fig:kbksast}
  Density dependence of $k_s^\ast$ and $k_b^\ast$ in the case of $r_p=0.8$ fm, $\nu = 2.0$ and $w=0.4$ GeV.  
   }
\end{figure*}

In the postsaturation region, i.e. $\rho_N > \rho_{sat}^\ast$, we introduce the depletion phase in the nucleon momentum distribution.  The boundary for the unoccupied bulk part, $k_b^\ast$, and the upper bound of the nucleon momentum, $k_s^\ast$, in SNM are evaluated by Eq.~(\ref{eq:rhoabove}), in which $k_b$ and $k_s$ are respectively replaced with $k_b^\ast$ and $k_s^\ast$, and by the modified boundary condition, Eq.~(\ref{eq:determination3}) with $k_b^\ast$ and $k_s^\ast$
\begin{equation}
g_Q^\ast(k_s^\ast) - \beta \alpha_\nu(k_b^\ast) g_Q^\ast(k_b^\ast) = 1  .  \label{eq:determination5}    
\end{equation}
For PNM, Eqs.~(\ref{eq:rhoaboven}) and (\ref{eq:determination4}) with $m^\ast$, $\epsilon^\ast$, and $a^\ast$ similarly generate the two momenta, $k_b^\ast$ and $k_s^\ast$.  Fig.~\ref{fig:kbksast} depicts the density dependence of $k_b^\ast$ and $k_s^\ast$.  In the calculation of the QQCM model, as discussed in the end of this section, we adjust the parameters in the regulator so as to reproduce the sound velocity inferred from the observed data of neutron stars and pressure determined by the experiment of heavy-ion collisions (HICs) at high energy.  We here take $\nu =2.0$ and $w=0.4$ GeV.  By choosing the large values of the parameters, the rapid rise of the momenta beyond $\rho_{sat}$ can be  completely controlled.  

Using Eqs.~(\ref{eq:nucleondist1}) and (\ref{eq:nucleondist2}) with $k_b^\ast$ and $k_s^\ast$, the total energy per nucleon in matter is given by Eq.~(\ref{eq:energydensity}) with 
\begin{align}
  \epsilon_N^\ast
  = &\frac{2}{\pi^{2}}  [ I(M_N^\ast, k_s^\ast) - \beta I(M_N^\ast, k_b^\ast) ]  
  \ \ \ \ \ \ \  \text{for SNM} ,  \label{eq:energydensitysnm} \\
  \epsilon_N^\ast
  = &\frac{1}{\pi^{2}}  [ I(M_N^\ast, k_s^\ast) - \beta^{\prime} I(M_N^\ast, k_b^\ast) ]   
   \ \ \ \ \ \ \ \ \text{for PNM}  .   \label{eq:energydensitypnm}
\end{align}
For the mean-field of the $\sigma$ meson, Eq.~(\ref{eq:sigmamf}) turns out to be 
\begin{align}
  \left(m_{\sigma}^{2}+g_{2}\bar{\sigma} \right) \bar{\sigma}
  &= g_\sigma G_{\sigma}(\bar{\sigma}) \left(\frac{2M_{N}^{\ast}}{\pi^{2}}\right) \left[ J(M_{N}^{\ast}, k_s^\ast) - \beta J(M_{N}^{\ast}, k_b^\ast) \right]  \ \ \ \ \ \ \ \text{for SNM}, 
  \label{eq:sigmamfsnm} \\
\left(m_{\sigma}^{2}+g_{2}\bar{\sigma} \right) \bar{\sigma}
  &= g_\sigma G_{\sigma}(\bar{\sigma}) \left(\frac{M_{N}^{\ast}}{\pi^{2}}\right) \left[ J(M_{N}^{\ast}, k_s^\ast) - \beta^{\prime} J(M_{N}^{\ast}, k_b^\ast) \right] \ \ \ \ \ \ \ \text{for PNM}
  \label{eq:sigmamfpnm}   . 
\end{align}
The mean-field of $\omega$ meson in SNM is calculated by Eq.~(\ref{eq:omegamf}), in which $\rho_N$ is given by Eq.~(\ref{eq:rhoabove}) with $k_b^\ast$ and $k_s^\ast$, instead of $k_b$ and $k_s$.  In case of PNM, the value of $\bar \omega$ is evaluated similarly, using Eq.~(\ref{eq:rhoaboven}) with $k_b^\ast$ and $k_s^\ast$.  In contrast, the mean-feild of $\rho$ meson in SNM vanishes because $\rho_3 = 0$, while, in PNM, it is calculated by Eq.~(\ref{eq:rhomf}), where $\rho_3$ is given by Eq.~(\ref{eq:rhoaboven}) with $k_b^\ast$ and $k_s^\ast$.  

The binding energy per nucleon,  pressure, the mean-field value for the $\sigma$ meson, the density dependences of the effective quark mass and of the effective nucleon mass, and the scalar polarizability in the QQMC model are respectively illustrated in Figs.~\ref{fig:Eb-Press}, \ref{fig:sigma-fields}, \ref{fig:Quark-Nucleon-mass} and \ref{fig:Gsigma}.   We note that the difference between those quantities in the QMC and QQMC models is small.  

\begin{figure*}[h!]
 \includegraphics[width=15.0cm,keepaspectratio,clip]{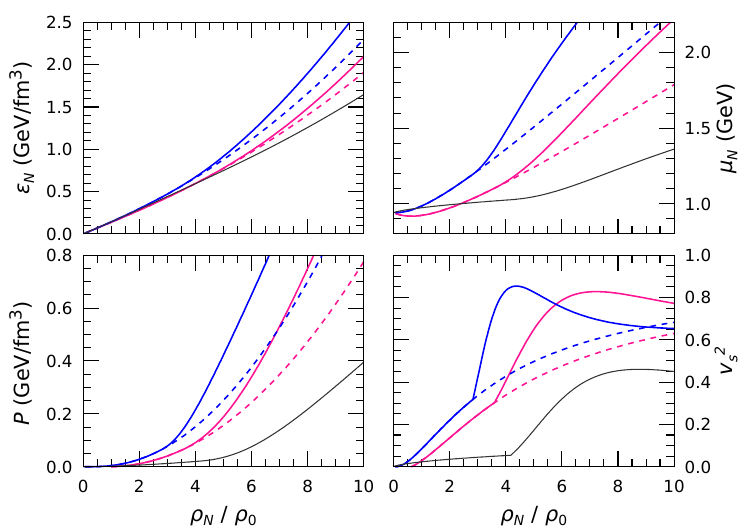}%
  \caption{\label{fig:qqmcresult06}
  Energy density, chemical potential, pressure and sound velocity in the case of $r_p=0.6$ fm, $\nu =2.0$, and $w=0.4$ GeV.  The red solid (dashed) lines denote the results of the QQMC (QMC) model in SNM, while the blue solid (dashed) lines denote those of the QQMC (QMC) model in PNM.  For comparison, we also show the results of the GQ model by the black lines. 
   }
\end{figure*}
\begin{figure*}[h!]
 \includegraphics[width=15.0cm,keepaspectratio,clip]{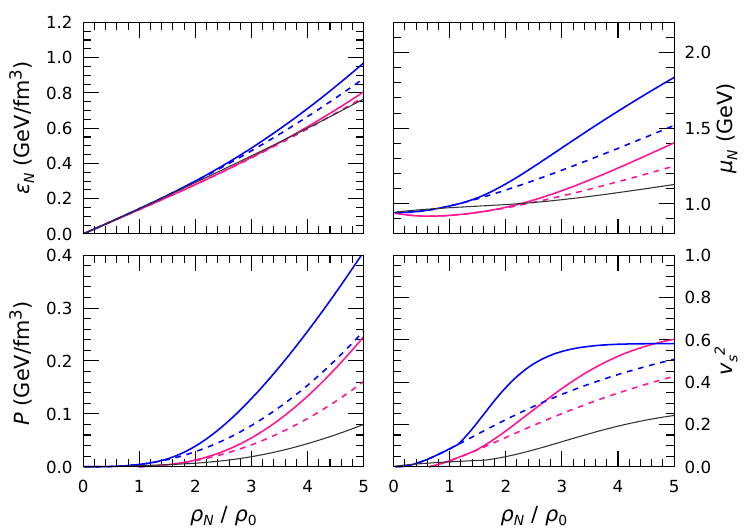}%
  \caption{\label{fig:qqmcresult08}
  Same as in Fig.~\ref{fig:qqmcresult06}, but for the case of $r_p=0.8$ fm.  
   }
\end{figure*}
Once we obtain the total energy density, $\epsilon_{tot}$, as a function of the nuclear density, $\rho_N$, we can calculate the chemical potential, pressure, and sound velocity numerically by 
\begin{equation}
\mu_N = \frac{\partial \epsilon_{tot}}{\partial \rho_N} , \ \ \ \ \ 
P = \mu_N \rho_N - \epsilon_{tot} , \ \ \ \ \ v_s^2 = \frac{\rho_N}{\mu_N}\frac{\partial \mu_N}{\partial \rho_N} = \frac{\rho_N}{\mu_N}\frac{\partial^2 \epsilon_{tot}}{\partial \rho_N^2}  .  \label{eq:observables}    
\end{equation}
Figs.~\ref{fig:qqmcresult06} and \ref{fig:qqmcresult08} present the results of the QQMC and QMC models in the case of $r_p = 0.6$ or $0.8$ fm, respectively.  For comparison, we add the results of the GQ model (see Sec.~\ref{subsec:correctedgaussianmodel}), which do not involve any nuclear interactions except for the quark Pauli blocking, to the figures.  It is seen that the nuclear interaction is very important to describe the properties of nuclear matter quantitatively, and that, as expected, the effect of Pauli blocking at the quark level enhances the chemical potential, pressure and the sound velocity beyond $\rho_{sat}$.  

As for the sound velocity, it strongly depends on the nucleon size.  The density dependence of $v_s^2$ is deduced from several neutron-star data by using the neural network model~\cite{PhysRevD.101.054016}, in which $v_s^2$ shows the characteristic feature, i.e. it increases rapidly around $\rho_N/\rho_0 \sim 4.0$ and reaches the maximum ($v_s^2 \sim 0.5 - 0.8$), and then it decreases gradually beyond that density.  However, such behavior can not be explained by the QMC model, in which $v_s^2$ grows up monotonously with increasing $\rho_N$.  In contrast, in the case of the QQMC model with $r_p=0.6$ fm (see Fig.~\ref{fig:qqmcresult06}), $v_s^2$ in PNM reaches the maximum, at which its value exceeds $0.8$, around $\rho_N/\rho_0 \sim 4.0$, and then decreases gradually.  Because a neutron star is composed of neutrons with a fraction of protons, it can be expected that the maximum value of $v_s^2$ in neutron-star matter turns out to be slightly smaller than the value in PNM.  Thus, the sound velocity calculated by the QQMC model is consistent with the result of $v_s^2$ inferred from the observed neutron-star data. 

On the other hand, in Refs.~\cite{Brandes_2024, PhysRevD.111.034005}, the chemical potential, pressure, and sound velocity are extracted from Bayesian inference analysis of the observed data of neutron stars.  Then, the analysis tells us that the sound velocity behaves differently than in Ref.~\cite{PhysRevD.101.054016}, and that it is enhanced rapidly around $\rho_N/\rho_0 \sim 2.0-3.0$, reaches the maximum ($v_s^2 \sim 0.6$) and is saturated beyond $\rho_N/\rho_0 \sim 3.0$.  This is very close to the present result with $r_p=0.8$ fm in PNM (see Fig.~\ref{fig:qqmcresult08}).  

Therefore, in the QQMC model, the value of $r_p$ should lie within the range of $0.6 - 0.8$ fm.  We here note that the behavior of $v_s^2$ observed in neutron stars can be explained by the OMEG family~\cite{10.3389/fphy.2024.1531475, MIYATSU2023138013}, particularly the set of OMEG1, in relativistic mean-field theory as well.

\begin{figure*}[h!]
 \includegraphics[width=12.0cm,keepaspectratio,clip]{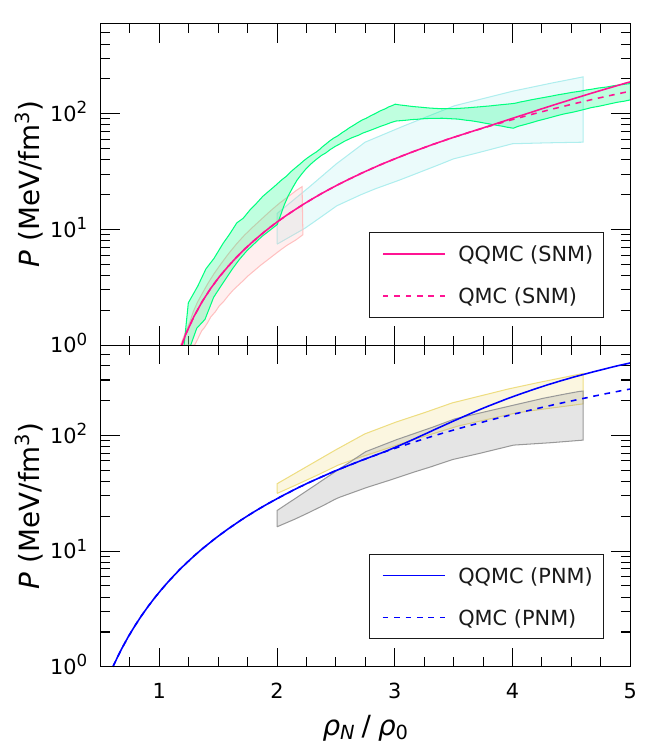}%
  \caption{\label{fig:qqmcP06}
  Pressure in SNM or PNM as a function of $\rho_N/\rho_0$.  We take $r_p=0.6$ fm, $\nu =2.0$   and $w=0.4$ GeV.  In SNM, the green and light blue areas are for the flow data of HICs~\cite{doi:10.1126/science.1078070, PhysRevC.108.034908}, while the pinkish  area is for the kaon production data~\cite{FUCHS20061, LYNCH2009427}.  In PNM, the yellow and gray areas are for the flow data of HICs~\cite{doi:10.1126/science.1078070}.  
   }
\end{figure*}
\begin{figure*}[h!]
 \includegraphics[width=12.0cm,keepaspectratio,clip]{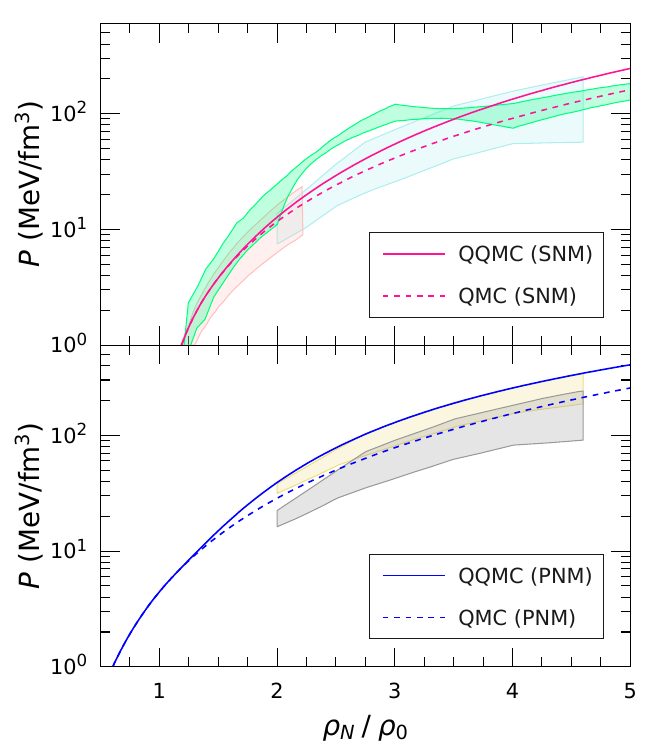}%
  \caption{\label{fig:qqmcP08}
  Same as in Fig.~\ref{fig:qqmcP06} but for $r_p=0.8$ fm, $\nu =2.0$ and $w=0.4$ GeV.  
   }
\end{figure*}
Furthermore, we can compare the present results with the experimental data of pressure extracted from HICs at high energy.  Figs.~\ref{fig:qqmcP06} and \ref{fig:qqmcP08} show such comparisons.  In the figures, the shaded areas represent the constraints from elliptical flow data~\cite{doi:10.1126/science.1078070, PhysRevC.108.034908} and kaon production  data~\cite{FUCHS20061, LYNCH2009427}.  
In both cases of $r_p = 0.6$ and $0.8$ fm, pressure calculated by the QMC or QQMC model is consistent with the experimental constraints.  We note that pressure at high density is mainly  generated by the repulsive force due to the exchanges of $\omega$ and $\rho$ mesons, and that the effect of quarkyonic phase contributes to pressure additionally.  

\clearpage

\newpage
\section{Summary and discussion} \label{sec:concandsum}

We have constructed a novel, practical nuclear model based on the quark degrees of freedom, namely the quarkyonic quark-meson coupling (QQMC) model, in which not only the variation of nucleon structure but also the effect of Pauli blocking at the quark level are taken into account.  In the QQMC model, a relativistic, gaussian quark wavefunction is adopted to describe the structure of nucleon.  

We summarize the present study as follows: 
\begin{enumerate}
\item The quark saturation density, $\rho_{sat}$, is very sensitive to the quark wavefunction. In the present calculation, we have chosen a relativistic gaussian wavefunction.  Due to the lower component of the relativistic wavefunction, the value of $\rho_{sat}$ turns out to be higher than that in the NR case.  Furthermore, it is also important to consider the c.m. correction to the estimate of a nucleon radius properly.  
For example, in the case of SNM in the ideal Fermi gas with $r_p = 0.8$ fm, we found $\rho_{sat} = 1.63 \rho_0$, which is higher than in the NR case by about $28\%$.  
In contrast, when the nuclear interaction is included, the quark saturation density is reduced because the nucleon size is swollen in matter.  In case of $r_p = 0.8$ fm in SNM, we found $\rho_{sat}^\ast = 1.46 \rho_0$, which is lower than the value of $1.63$ by about $10\%$.  Thus, the quark saturation density in SNM becomes eventually larger by about $15\%$, compared with the value in the NR case.  We note that $\rho_{sat}^\ast$ in PNM is roughly $22\%$ lower than that in SNM.  In the present calculation, 
$\rho_{sat}^\ast$ in SNM is {\it never} below the nuclear saturation density, $\rho_0$, which is  consistent with the present status of nuclear experiments.  

\item The quarkyonic phase is characterized by two momenta, $k_b$ and $k_s$, where the former defines the under-occupied bulk part at low momentum and the latter gives the upper bound of the shell structure at high momentum.  In the present paper, to determine those two momenta, we have used the same boundary condition as in the IdylliQ model.  Then, using the gaussian quark wavefunction, we have first evaluated the energy density, chemical potential, pressure and sound velocity within the ideal Fermi gas picture.  In this na{\" i}ve GQ model, the chemical potential and pressure are discontinuous and the sound velocity diverges at $\rho_{sat}$.  As a minimal correction to remedy this singular behavior, we have adopted a regulator which smears the sharp Fermi surface, $\theta(k_b-k)$, in the nucleon momentum distribution.  Then, optimizimg the regulator, we have found that the chemical potential and pressure are smoothed at $\rho_{sat}$, and that the sound velocity becomes continuous at $\rho_{sat}$.  

\item We have next combined the GQ model and the QMC model in order to introduce the nuclear interaction -- it is called the QQMC model.  The QQMC model is a nuclear model based on the quark degrees of freedom, and it is available to use over the wide range of nuclear density from low density to the crossover region where the transition from baryonic to quark matter occurs.  The new model includes the effect of Pauli blocking at the quark level as well as the effect of scalar polarizability of nucleon in nuclear medium.  Therefore, the quarks feel the scalar field condensed in matter, which induces many-body forces.  Then, we have calculated the physical quantities like the energy density, chemical potential, pressure and sound velocity, which include the effect of nuclear interaction.  

\item We emphasize that the inclusion of nuclear interaction in the quarkyonic model is quite important to consider the physical quantities quantitatively.  In fact, the QQMC model can provide the sound velocity which is consistent with that inferred from the observed neutron-star data by using the neural network model or Bayesian inference analysis.  Furthermore, pressure in PNM or SNM calculated with the QQMC model is within the range deduced from the experiments of HICs at high energy.  As the quark saturation density heavily relies on the size of nucleon, various physical quantities also depend on it accordingly.  From the consideration on the sound velocity, the choice of $r_p = 0.6 - 0.8$ fm seems most suitable for describing dense nuclear matter in the QQMC model.  
\end{enumerate}

There are still some problems left.  The first one concerns the boundary condition to determine the two momenta, $k_b$ and $k_s$.  Although in the present study we have assumed the condition which is the same as in the IdylliQ model, we need further investigation to make sure whether such condition is {\it sufficient} or not.  At the same time, it is also necessary to find a general rule for constructing the quark momentum distribution, $f_Q(q)$, in the postsaturation region, which satisfies Fermi statistics and the minimization of energy.  Further rigorous investigation on this issue is urgent.  The second one relates to the singular behavior of physical quantities around the quark saturation density.  In the present calculation, we have used an {\it ad hoc} regulator to remove it.  However, such singular behavior should be basically resolved by the fundamental dynamics, i.e. QCD -- for example, see Refs.~\cite{PhysRevD.105.076001, 4ywp-752m}.  Those are the future issues.  

Finally, it may be possible to include temperature in the QQMC model and to apply the model to EoSs for neutron stars with hyperons~\cite{txbp-t8vm, PhysRevC.111.044914}.

\newpage
\section*{Acknowledgments}

K.~S. thanks Yuki Fujimoto for valuable discussions on Quarkyonic matter. 
This work was supported by the Basic Science Research Program through the National Research Foundation of Korea (NRF) under Grant Nos. RS-2026-25471436,  RS-2025-16071941, RS-2023-00242196, and RS-2021-NR060129.  

\appendix

\section{Sommerfeld expansion for a gaussian quarkyonic model}  \label{app:sommerfeld}

In nuclear physics, for example, the Fermi function
\begin{equation}
 f(z) = \frac{1}{1 + e^z}  ,  \ \ \ \ \ \ \  z= \frac{r-R}{\delta}       \label{eq:fermifunction}   
\end{equation}
with $R$ the half-density radius and $\delta$ the surface thickness is often used to describe a shape for the radial density distribution of a nucleus.  However, it is more convenient to use the symmetrized Fermi function defined by~\cite{DWLSprung1997}
\begin{align}
 f_s(r, R) &\equiv f((r-R)/\delta) + f(-(r+R)/\delta) -1    \label{eq:symfermifunction}   \\
 &=  f((r-R)/\delta) - f((r+R)/\delta) = \frac{\sinh(R/\delta)}{\cosh(r/\delta) + \cosh(R/\delta)}  ,  \nonumber 
\end{align}
where $f_s(r, R) = f_s(-r, R)$, because it can avoid some difficulties appearing in actual calculations.  Here, instead of the usual Fermi function, $f(z)$, we consider minimal corrections to the GQ model by using the symmetrized Fermi function, $f_s(r, R)$.   

First, we formulate Sommerfeld expansion generally.  We consider the integral 
\begin{equation}
K[h, p) = \int_0^\infty dk \, h(k) f_s(k, p) ,        \label{eq:intkhp}   
\end{equation}
with $h(k)$ any smooth function.  Defining 
\begin{equation}
H(k) \equiv \int_0^k dk^\prime \, h(k^\prime) ,  \ \ \ \ \ \ \ H(0) = 0  ,        \label{eq:hh}   
\end{equation}
we integrate Eq.~(\ref{eq:intkhp}) by parts, so that 
\begin{align}
K[h, p) &= - \int_0^\infty dk \, H(k) \frac{\partial}{\partial k} f_s(k, p) 
= - \int_0^\infty dk \, H(k) \frac{\partial}{\partial k} \left[ f(x-y) - f(x+y) \right] ,      \nonumber \\
&= - \int_0^\infty dk \, H(k) \left[ \frac{\partial u}{\partial k} f^\prime(u) - 
\frac{\partial v}{\partial k} f^\prime(v) \right] = - \frac{1}{\delta} \int_0^\infty dk \, H(k) \left[ f^\prime(u) - f^\prime(v) \right]  , \label{eq:intkhp2}   
\end{align}
with $x=k/\delta$, $y=p/\delta$, $u=x-y=(k-p)/\delta$ and $v=x+y=(k+p)/\delta$.  Then, $H(k)$ is expended around $k=p$ up to ${\cal O}((k-p)^2)$
\begin{equation}
H(k) \simeq H(p) + \delta u h(p) + \frac{1}{2} \delta^2 u^2 h^{\prime}(p) ,  \label{eq:hexp}   
\end{equation}
with $h^{\prime}(p) = \frac{d}{dk} h(k)|_{k=p}$.  Thus, we find 
\begin{align}
K[h, p) &\simeq - \frac{1}{\delta} H(p) \int_0^\infty dk \, \left[ f^\prime(u) - f^\prime(v) \right] 
-  h(p) \int_0^\infty dk \, u \left[ f^\prime(u) - f^\prime(v) \right]   \nonumber   \\
&- \frac{\delta}{2} h^\prime(p) \int_0^\infty dk \, u^2 \left[ f^\prime(u) - f^\prime(v) \right] , 
\label{eq:intkhp3}
\end{align}
where the integrals in the right-hand side are evaluated as
\begin{align}
\chi_0^-(y) &\equiv - \frac{1}{\delta} \int_0^\infty dk \, \left[ f^\prime(u) - f^\prime(v) \right] 
= f_s(0, y) = \tanh \left( \frac{y}{2} \right) ,   \label{eq:chi0-}  \\
\chi_1^-(y) &\equiv  - \frac{1}{\delta} \int_0^\infty dk \, u \left[ f^\prime(u) - f^\prime(v) \right]  
= \frac{2y}{1+e^y}  , \label{eq:chi1-}   \\
\chi_2^-(y) &\equiv  - \frac{1}{\delta} \int_0^\infty dk \, u^2 \left[ f^\prime(u) - f^\prime(v) \right] = \frac{\pi^2}{3} S_2(y) + 4y \log(1+e^{-y}) ,   \label{eq:chi2-}
\end{align}
with 
\begin{equation}
S_2(y) \equiv \frac{12}{\pi^2} \int_0^{y/2} dt \, \frac{t^2}{\cosh^2 t} 
\simeq \frac{y^3}{y^3 + (19y^2 -17y + 40)e^{-0.6y}}  \ \ \ \ \ (0 \leq S_2(y) \leq 1)  .   \label{eq:s2}
\end{equation}
Note that, because of the symmetrized Fermi function, the integrals (except $S_2$) can be performed without any approximation.  Eventually, up to ${\cal O}(\delta^2)$, Sommerfeld expansion gives 
\begin{equation}
K[h, p) \simeq \chi_0^-(y) H(p) + \delta \chi_1^-(y) h(p) + \frac{1}{2} \delta^2 \chi_2^-(y) h^\prime(p)  . 
\label{eq:intkhp4}
\end{equation}

Similarly, we can find the first and second derivatives of $K[h, p)$ up to ${\cal O}(\delta^2)$ as
\begin{equation}
\frac{\partial}{\partial p} K[h, p) \simeq h(p) + \delta \chi_1^+(y) h^\prime(p) + \frac{1}{2} \delta^2 \chi_2^+(y) h^{\prime \prime}(p)  ,    \label{eq:derintkhp1}
\end{equation}
with
\begin{align}
\chi_1^+(y) &\equiv  - \frac{1}{\delta} \int_0^\infty dk \, u \left[ f^\prime(u) + f^\prime(v) \right]  
= 2 \log(1+e^{-y})  ,  \label{eq:chi1+}   \\
\chi_2^+(y) &\equiv  - \frac{1}{\delta} \int_0^\infty dk \, u^2 \left[ f^\prime(u) + f^\prime(v) \right] = \frac{\pi^2}{3} - 4y \log(1+e^{-y}) ,   \label{eq:chi2+}
\end{align}
and 
\begin{equation}
\frac{\partial^2}{\partial p^2} K[h, p) \simeq \chi_0^-(y) h^\prime(p) + \delta \chi_1^-(y) h^{\prime \prime}(p) + \frac{1}{2} \delta^2 \chi_2^-(y) h^{\prime \prime \prime}(p)  .    \label{eq:der2intkhp1}
\end{equation}

Next, using the formula, Eqs.~(\ref{eq:intkhp4}), (\ref{eq:derintkhp1}) and (\ref{eq:der2intkhp1}), we consider the minimal corrections to the GQ model, that is, we want to modify the first and second derivatives, $\partial k_b/\partial k_s$ and $\partial^2 k_b/\partial k_s^2$, in the na{\" i}ve GQ model so as to remove the singular behavior at the quark saturation density.  Here, we focus on the case of SNM only.  The PNM case can be handled similarly as well.  

\begin{figure*}[h!]
 \includegraphics[width=12.0cm,keepaspectratio,clip]{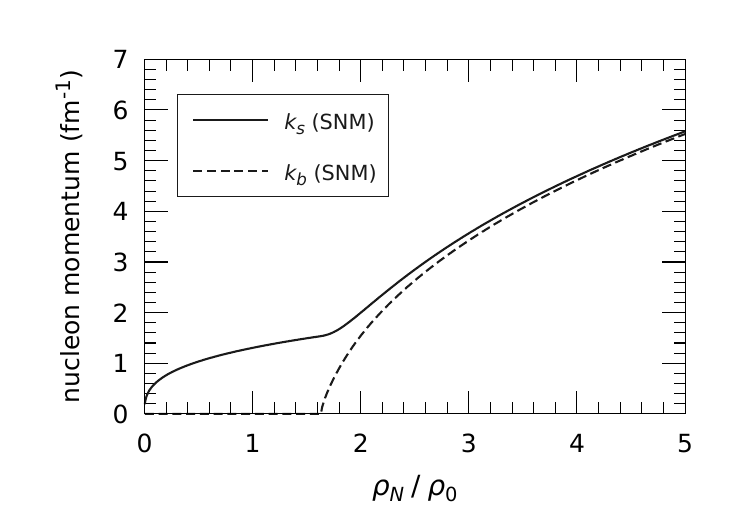}%
  \caption{\label{fig:correctkbks2}
   Momenta, $k_s$ and $k_b$, in Sommerfeld expansion.  We choose $r_p = 0.8$ fm and $\delta = 35$ MeV.   
   }
\end{figure*}

In the boundary conditions, Eq.~(\ref{eq:determination1}), we replace $g_{Q}(k_b)$ with $K[h, p)$.  In Eq.~(\ref{eq:intkhp4}), in the limit $\delta \to 0$, $K[h, p)$ approaches $H(p)$, so that we set $p \to k_b$, $H \to g_Q(k_b)$ and $h \to \phi$ (see Eq.~(\ref{eq:dergq})).  Therefore, the boundary condition with the corrections reads
\begin{equation}
g_Q(k_s) - \beta K[\phi, k_b) = 1   ,     \label{eq:moddetermination1}    
\end{equation}
and 
\begin{equation}
K[\phi, k_b) \simeq \chi_0^-(k_b/\delta) g_Q(k_b) + {\bar \delta} \chi_1^-(k_b/\delta) \phi(k_b) + \frac{1}{2} {\bar \delta}^2 \chi_2^-(k_b/\delta) \phi^\prime(k_b)  ,   \label{eq:intkhp5}
\end{equation}
where ${\bar \delta} = a\delta/N_c$ is dimensionless and $\phi^\prime$ is given by Eq.~(\ref{eq:derphi}).  Fig.~\ref{fig:correctkbks2} depicts the momenta, $k_b$ and $k_s$, calculated with Sommerfeld expansion, Eq.~(\ref{eq:moddetermination1}).  We can see that the rapid rise of the momenta appeared in the na{\" i}ve GQ model is softened at $\rho_{sat}$.  

Furthermore, the first derivative of $K[\phi, k_b)$ with respect to ${\bar k}_b$ is given by Eq.~(\ref{eq:derintkhp1}) 
\begin{equation}
\frac{\partial K[\phi, k_b)}{\partial {\bar k}_b} \simeq \phi(k_b) + {\bar \delta} \chi_1^+(k_b/\delta) \phi^\prime(k_b) + \frac{1}{2} {\bar \delta}^2 \chi_2^+(k_b/\delta) \phi^{\prime \prime}(k_b)  .   \label{eq:derintkhp2}
\end{equation}
Thus, we obtain the first derivative, $\partial k_b/\partial k_s$, as
\begin{align}
\frac{\partial k_b}{\partial k_s}  &= \frac{\partial {\bar k}_b}{\partial {\bar k}_s}  \nonumber \\
&= \frac{\phi(k_s)}{\beta [\phi(k_b) + {\bar \delta} \chi_1^+(k_b/\delta) \phi^\prime(k_b) + \frac{1}{2} {\bar \delta}^2 \chi_2^+(k_b/\delta) \phi^{\prime \prime}(k_b)]} ,   \label{eq:derintkhp5}
\end{align}
with
\begin{equation}
\phi^{\prime \prime}(k) = \frac{N_c^3}{\sqrt{\pi}} 8 [r_1(k) - 5 r_2(k) {\bar k}^2 + 2 r_3(k) {\bar k}^4] e^{- {\bar k}^2}  .  \label{eq:derderphi}
\end{equation}
We notice that, as $k_b \to 0$, the denominator in Eq.~(\ref{eq:derintkhp5}) is ${\cal O}(1)$ and thus the singular behavior at $\rho_{sat}$ vanishes.  

To calculate the sound velocity, we need the second derivative, $\partial^2 k_b/\partial k_s^2$, which should be modified as well.  Using Eqs.~(\ref{eq:der2intkhp1}), (\ref{eq:moddetermination1}), and (\ref{eq:derintkhp5}), we find 
\begin{equation}
\frac{\partial^2 {\bar k}_b}{\partial {\bar k}_s^2} = \left( \frac{N_c}{a} \right) \frac{\partial^2 k_b}{\partial k_s^2} = \frac{\phi^\prime(k_s) -\beta \left( \frac{\partial {\bar k}_b}{\partial {\bar k}_s} \right)^2 \frac{\partial^2}{\partial {\bar k}_b^2} K[\phi, k_b)}{\beta [\phi(k_b) + {\bar \delta} \chi_1^+(k_b/\delta) \phi^\prime(k_b) + \frac{1}{2} {\bar \delta}^2 \chi_2^+(k_b/\delta) \phi^{\prime \prime}(k_b)]}   ,   \label{eq:der2intkhp2}
\end{equation}
with
\begin{equation}
\frac{\partial^2}{\partial {\bar k}_b^2} K[\phi, k_b) \simeq \chi_0^-(k_b/\delta) \phi^\prime(k_b) + {\bar \delta} \chi_1^-(k_b/\delta) \phi^{\prime \prime}(k_b) + \frac{1}{2} {\bar \delta}^2 \chi_2^-(k_b/\delta) \phi^{\prime \prime \prime}(k_b)  ,     \label{eq:der2intkhp3}
\end{equation}
and 
\begin{equation}
\phi^{\prime \prime \prime}(k) = \frac{N_c^3}{\sqrt{\pi}} (-16){\bar k} [6r_2(k) - 9 r_3(k) {\bar k}^2 + 2 r_4(k) {\bar k}^4] e^{- {\bar k}^2}  .  \label{eq:derderderphi}
\end{equation}
We again notice that the denominator in Eq.~(\ref{eq:der2intkhp2}) is ${\cal O}(1)$ as $k_b \to 0$.  

\begin{figure*}[h!]
 \includegraphics[width=15.0cm,keepaspectratio,clip]{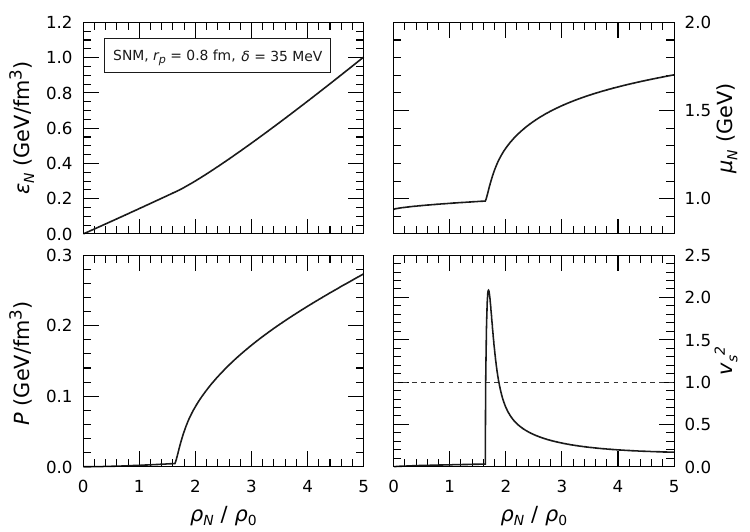}%
  \caption{\label{fig:revsnmresult}
   Energy density, chemical potential, pressure, and sound velocity in Sommerfeld expansion. 
   }
\end{figure*}
Now, all we need is prepared.  Along the line of Section\,\ref{subsec:results1}, we can calculate $\rho_N$, $\epsilon_N$, $\mu_N$, $P$, and $v_s^2$ by using Eqs.~(\ref{eq:derintkhp5}) and (\ref{eq:der2intkhp2}), instead of Eqs.~(\ref{eq:derkbks}) and (\ref{eq:derkbks2}).  Fig.~\ref{fig:revsnmresult} illustrates the results.  We take $\delta = 35$ MeV and $r_p=0.8$ fm.  All the calculated quantities are continuous at $\rho_{sat}$.   Furthermore, the sound velocity does not diverge.  However, it overshoots the upper limit. 
A larger value of $\delta$ may, of course, suppress more the sound velocity around $\rho_{sat}$, but it turns out to be negative at very high density.  Therefore, the present expansion up to ${\cal O}(\delta^2)$ may be insufficient, and we should include higher order contributions beyond ${\cal O}(\delta^2)$ to obtain a satisfactory result.  It may be very intricate.

\newpage

\bibliographystyle{apsrev4-2.bst}
\bibliography{qqmc.bib}

\end{document}